\documentclass[journal]{IEEEtran}
\usepackage{amsmath,amsfonts}
\usepackage{algorithmic}
\usepackage{algorithm}
\usepackage{array}
\usepackage[caption=false,font=normalsize,labelfont=sf,textfont=sf]{subfig}
\usepackage{textcomp}
\usepackage{stfloats}
\usepackage{url}
\usepackage{verbatim}
\usepackage{multirow}
\usepackage{graphicx}
\usepackage{hyperref}
\usepackage{cite}
\usepackage{xcolor}
\usepackage{tabularx}
\usepackage[commandnameprefix=ifneeded,defaultcolor=black]{changes}

\begin{document}

\author{Bang Zeng, Student Member, IEEE, Ming Li\IEEEauthorrefmark{1}, Senior Member, IEEE
\IEEEcompsocitemizethanks{
	\IEEEcompsocthanksitem Bang~Zeng and Ming~Li are with the School of Computer Science, Wuhan University, Wuhan 430072, China, and also with Suzhou Municipal Key Laboratory of Multimodal Intelligent Systems, Digital Innovation Research Center, Duke Kunshan University, Kunshan 215316, China (e-mail: bangzeng@whu.edu.cn; ming.li369@dukekunshan.edu.cn).}
\thanks{\IEEEauthorrefmark{1} Corresponding author.}}


\title{USEF-TSE: Universal Speaker Embedding Free Target Speaker Extraction}
\maketitle

\begin{abstract}
Target speaker extraction aims to separate the voice of a specific speaker from mixed speech. Traditionally, this process has relied on extracting a speaker embedding from a reference speech, in which a speaker recognition model is required. However, identifying an appropriate speaker recognition model can be challenging, and using the target speaker embedding as reference information may not be optimal for target speaker extraction tasks. This paper introduces a Universal Speaker Embedding-Free Target Speaker Extraction (USEF-TSE) framework that operates without relying on speaker embeddings. USEF-TSE utilizes a multi-head cross-attention mechanism as a frame-level target speaker feature extractor. This innovative approach allows mainstream speaker extraction solutions to bypass the dependency on speaker recognition models and better leverage the information available in the enrollment speech, including speaker characteristics and contextual details. Additionally, USEF-TSE can seamlessly integrate with other time-domain or time-frequency domain speech separation models to achieve effective speaker extraction. Experimental results show that our proposed method achieves state-of-the-art (SOTA) performance in terms of Scale-Invariant Signal-to-Distortion Ratio (SI-SDR) on the WSJ0-2mix, WHAM!, and WHAMR! datasets, which are standard benchmarks for monaural anechoic, noisy and noisy-reverberant two-speaker speech separation and speaker extraction. \added{The results on the LibriMix and the blind test set of the ICASSP 2023 DNS Challenge demonstrate that the model performs well on more diverse and out-of-domain data.} For access to the source code, please visit: \url{https://github.com/ZBang/USEF-TSE} 
\end{abstract}


\section{Introduction}
\IEEEPARstart{H}{umans} have the remarkable ability to selectively focus on a specific speech signal in a noisy environment, a skill often referred to as the cocktail party effect~\cite{Cherry1953SomeEO, bronkhorst2000cocktail}. Speech separation has emerged as a practical solution to address this challenge. It plays a vital role as a preliminary step for speech signal processing technologies, including speech recognition, speaker recognition, and speaker diarization~\cite{raj2021integration,delcroix2018single,xu2021target,Rao2019TargetSE,park2022review}.

Traditional speech separation algorithms, such as Non-negative Matrix Factorization (NMF)~\cite{schmidt2006single,cichocki2006new} and Computational Auditory Scene Analysis (CASA)~\cite{lyon1983computational,wang2006computational,hu2007auditory}, typically use spectro-temporal masking to separate each speaker’s voice from the mixed speech. These methods have significantly advanced speech separation technology. With the rapid advancement of deep learning, new methods that employ deep neural networks to estimate mask matrices have emerged, offering improved performance and adaptability.

Deep neural network-based speech separation algorithms, such as Deep Clustering~\cite{hershey2016deep,isik2016single,wang2018multi}, Deep Attractor Network (DANet)~\cite{chen2017deep,wang2018supervised,luo2018speaker}, and Permutation Invariant Training (PIT)~\cite{yu2017permutation,kolbaek2017multitalker}, have remarkably enhanced separation performance. Moreover, PIT has effectively addressed the label permutation problem. However, the Time-domain Audio Separation Networks (TasNet)~\cite{luo2018tasnet} highlights a significant limitation of time-frequency domain speech separation algorithms: their inability to accurately reconstruct the phase information of the clean speech. \added{Using only the magnitude spectrum in a frequency domain model can still perform well because the phase information is not critical in many tasks, such as in speaker recognition for the target speaker~\cite{rikhye2021personalized}. However, TasNet suggests that the effective reconstruction of phase information can enhance the performance limit of Speech Separation (SS) tasks optimized for human auditory perception.} TasNet addresses this issue  by using convolutions and deconvolutions in place of the short-time Fourier transform (STFT) and inverse STFT (iSTFT) for modeling and reconstructing target speech. Consequently, time-domain speech separation approaches~\cite{luo2020dual,tzinis2020sudo,luo2019conv,zeghidour2021wavesplit,li2021dual,chen2020dual,subakan2021attention,li2022efficient,rixen2022qdpn,yip2024spgm,zhao2023mossformer,zhao2024mossformer2} have been widely adopted. Various time-domain methods, such as Dual-Path RNN (DPRNN)~\cite{luo2020dual}, SepFormer~\cite{subakan2021attention} and Mossformer2~\cite{zhao2024mossformer2}, have significantly improved separation performance. Recently, the introduction of TF-GridNet~\cite{wang2023tf} has led to renewed excellence in time-frequency domain speech separation~\cite{rixen2022sfsrnet,yang2022tfpsnet,dang2022dpt}. Despite these advances, many speech separation methods still require prior knowledge of the number of speakers in the mixed speech. This prior knowledge is not always available in real-world applications, which poses challenges for applying these separation solutions in practical scenarios. 

\begin{figure*}[t!]
  \centering
  \includegraphics[width=0.75\linewidth]{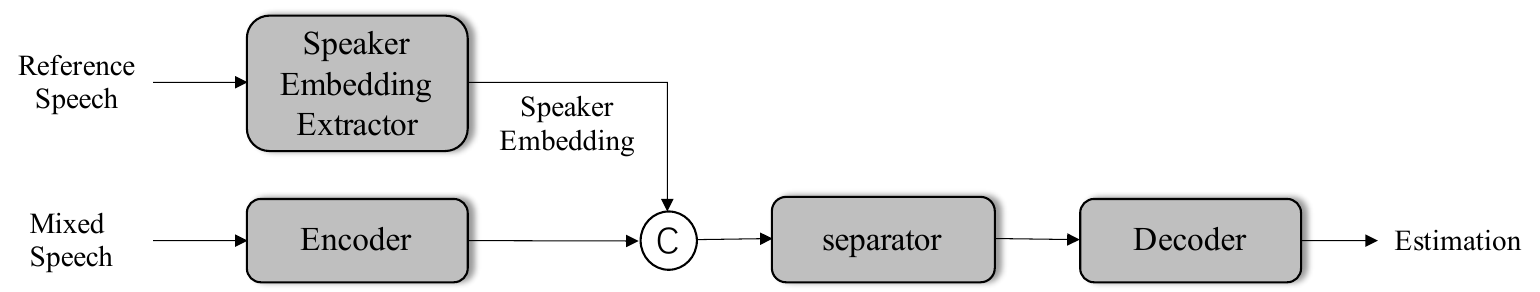}
  \caption{The diagram of a typical target speaker extraction method. The speaker embedding extractor is typically a pre-trained speaker recognition model. 'C' denotes the concatenation.}
  \label{fig:tytse}
\end{figure*}

Recently, Target Speaker Extraction (TSE)~\cite{wang19h_interspeech,vzmolikova2019speakerbeam,delcroix2020improving,hao2020unified,li20p_interspeech,Zhang2020XTaSNetRA,xu2020spex,ge2020spex+,wang2021neural,ge2021multi,liu2023x,hao2024x,elminshawi2022new} models have emerged as essential tools for addressing the limitations of traditional speech separation methods, especially in scenarios where the number of speakers in the audio mixture is unknown. By leveraging reference speech, these models can effectively isolate the target speaker’s voice from complex audio mixtures, making them highly applicable in real-world situations. This innovative approach not only improves the usability of speech separation technology but also expands its applicability in diverse and dynamic acoustic environments. Figure~\ref{fig:tytse} depicts a typical TSE framework. In the time-frequency (T-F) domain, the encoder and decoder represent the STFT and iSTFT operations, respectively. In the time domain, these components usually correspond to convolution and deconvolution operations. Traditional TSE models rely on a speaker embedding extractor to obtain the target speaker’s embedding from the reference speech. This extractor, such as a ResNet-based models~\cite{he2016deep,chung2020in,wan2018generalized,deng2019arcface}, can be pre-trained or jointly trained with the separator through multi-task learning. Building upon the framework illustrated in Figure~\ref{fig:tytse}, various techniques have been developed to enhance speaker extraction performance. However, these speaker embedding extraction models aim to maximize speaker recognition performance by training the embedding extractor. Consequently, these TSE models may not fully utilize the information available in the reference speech. As a result, approaches relying on speaker embeddings for the speaker extraction task may not be optimal. 

In our previous work, we are the first to propose a time-domain speaker embedding-free target speaker extraction model (SEF-Net)~\cite{zeng2023sef}. It offers a viable new solution for separating a target speaker’s voice without relying on speaker embeddings. In SEF-Net~\cite{zeng2023sef}, both the mixed speech and reference speech are processed using twin weights-sharing conformer~\cite{gulati20_interspeech} encoders. SEF-Net~\cite{zeng2023sef} then employs cross multi-head attention within the transformer~\cite{vaswani2017attention} decoder to implicitly leverage the speaker information in the reference speech’s conformer encoder outputs. SEF-Net~\cite{zeng2023sef} has achieved comparable performance to typical TSE models, highlighting the effectiveness of the speaker embedding-free framework for TSE tasks.

While SEF-Net~\cite{zeng2023sef} demonstrates significant potential in TSE tasks, it does have some limitations. Firstly, SEF-Net~\cite{zeng2023sef} requires either repetition or zero-padding to align the length of the reference speech with the mixed speech, which poses practical challenges. Secondly, SEF-Net~\cite{zeng2023sef} utilizes a multi-head cross-attention mechanism within both intra-block and inter-block modules to facilitate interaction between the features of mixed and reference speech. However, this approach does not fully exploit global information, potentially constraining the model’s performance improvements. Thirdly, SEF-Net~\cite{zeng2023sef} has yet to be tested on data with noise and reverberation, leaving its effectiveness in natural, noisy environments unverified. Lastly, unlike speaker other embedding-based TSE methods, SEF-Net~\cite{zeng2023sef} is not a universal framework. It cannot be easily integrated with arbitrary time-domain or time-frequency domain speech separation networks for TSE tasks.

To address the aforementioned issues, this paper proposes a Universal Speaker Embedding-Free Target Speaker Extraction (USEF-TSE) framework for monaural TSE based on SEF-Net~\cite{zeng2023sef}. USEF-TSE employs a unified encoder to process both the reference and mixed speech. The encoding of the mixed speech is utilized to query the encoding of the reference speech through a Cross Multi-Head Attention (CMHA) module. The output of the CMHA module is a frame-level feature that maintains the same length as the encoded features of the mixed speech. This frame-level feature represents the target speaker’s attributes and is fused with the encoding of the mixed speech. The resulting fused feature is then fed into a separator to model the target speaker’s voice components. \added{The differences between USEF-TSE and SEF-Net are shown in the Table~\ref{tab:diff}.}

%
\begin{table}[]
\caption{\added{The differences between USEF-TSE and SEF-Net.}}
  \label{tab:diff}
  \centering
\resizebox{0.48\textwidth}{!}{\Huge
\begin{tabular}{c|c|c}
\hline
\textbf{\added{Component}} & \textbf{\added{USEF-TSE}}                                                                                                                                                & \textbf{\added{SEF-Net}}                                                                                                                                            \\ \hline
\added{Encoder}       & \added{STFT or Conv1d}                                                                                                                                          & \added{Conv1d}                                                                                                                                             \\ \hline
\begin{tabular}[c]{@{}c@{}} \added{Fusion} \\ \added{Module} \end{tabular} & \begin{tabular}[c]{@{}c@{}}\added{Before the separator, the queries} \\ \added{come from the mixture, while}\\ \added{the keys and values} \\ \added{come from the reference speech.} \end{tabular} & \begin{tabular}[c]{@{}c@{}}\added{In the separator, the queries}\\ \added{come from the reference speech,}\\ \added{while the keys and values} \\ \added{come from the mixture.}\end{tabular} \\ \hline
\added{Separator}     & \begin{tabular}[c]{@{}c@{}}\added{The backbone of SS models or}\\ \added{network layers such as LSTM}\end{tabular}                                                          & \added{Transformer}                                                                                                                                        \\ \hline
\added{Decoder}       & \added{iSTFT or Transposed Conv1d}                                                                                                                               & \added{Transposed Conv1d}                                                                                                                                   \\ \hline
\added{Input}         & \begin{tabular}[c]{@{}c@{}}\added{No specific constraints}\\ \added{or limitations}\end{tabular}                                                                        & \begin{tabular}[c]{@{}c@{}}\added{The length of the reference }\\ \added{needs to be consistent with}\\ \added{the length of the mixture}\end{tabular}                               \\ \hline
\end{tabular}
}
\end{table}
This paper builds upon and extends our previous work on speaker embedding-free approaches for monaural TSE. The key contributions of this article are summarized as follows:
\begin{itemize}
\item{\textbf{Improved:} We address the limitation of SEF-Net, which requires the lengths of the reference and mixed speech to be the same. Additionally, we have altered the interaction position of the reference and mixed speech, enhancing the model’s ability to learn global information more effectively. Experimental results demonstrate that these modifications significantly improve the model’s performance.}
\item{\textbf{Universal:} We introduce USEF-Net, a universal speaker embedding-free TSE framework. USEF-Net can seamlessly integrate with most deep neural network-based time-domain or time-frequency domain speech separation models for TSE. To validate the effectiveness of the USEF-TSE, we propose time and T-F domain TSE models, namely USEF-SepFormer and USEF-TFGridNet. Experimental results show that our proposed methods achieve state-of-the-art (SOTA) performance on the WSJ0-2mix~\cite{hershey2016deep} datasets.}
\item{\textbf{Robust:} To assess the real-world performance of the USEF-TSE framework, we evaluated it on the WHAM!~\cite{Wichern2019WHAM} and WHAMR!~\cite{Maciejewski2020WHAMR} datasets. \added{Moreover, we use LibriMix and blind test set of ICASSP 2023 DNS Challenge to evaluate the stability and generalizability of USEF-TSE framework, respectively.}}
\end{itemize}
\begin{figure*}[t!]
  \centering
  \includegraphics[width=0.80\linewidth]{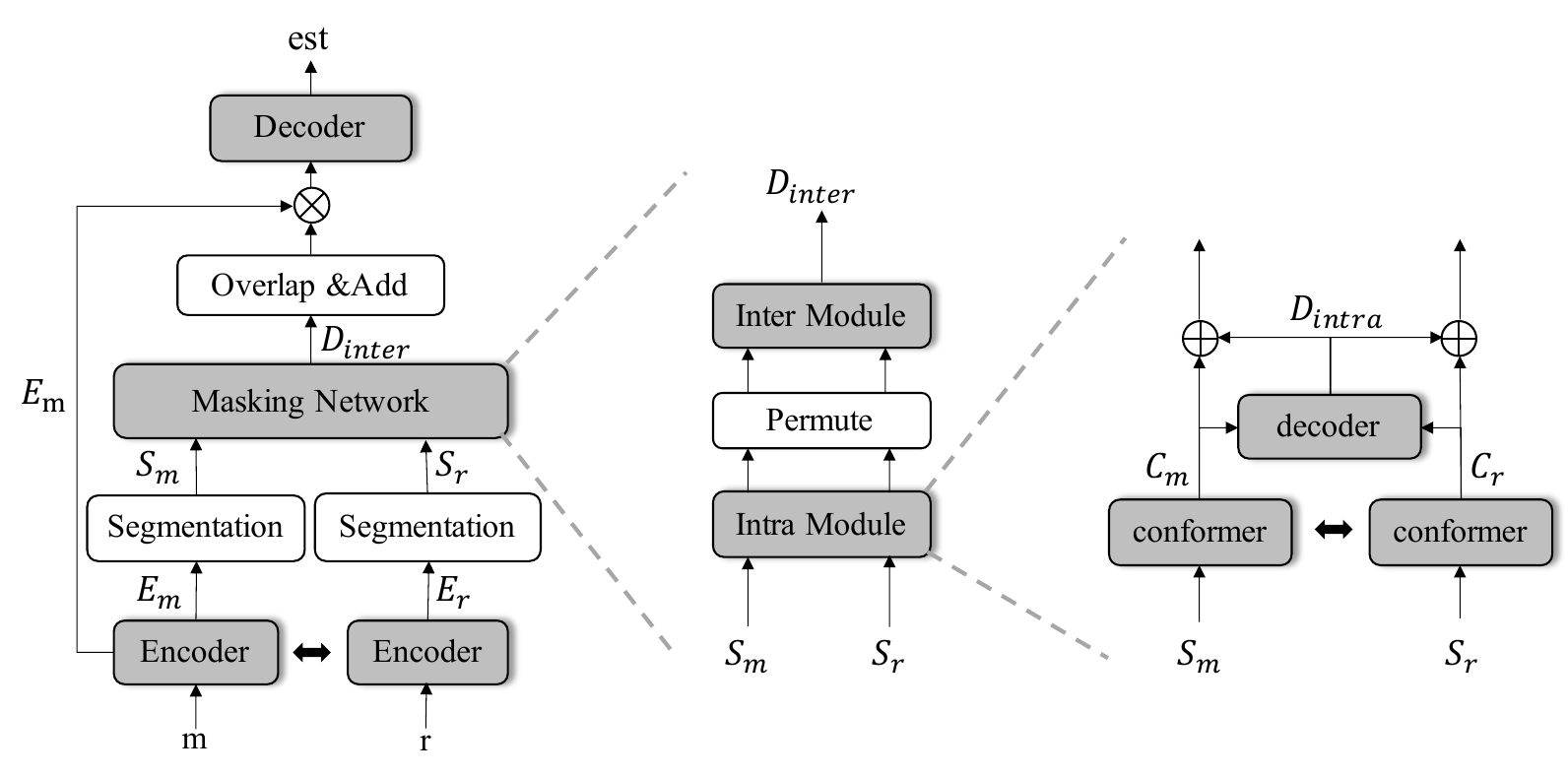}
  \caption{The diagram of the SEF-Net. ‘m’ and ‘r’ denote the mixed speech and \added{reference} speech, respectively. \added{‘\(\textbf{E}_{m}\)’ and ‘\(\textbf{E}_{r}\)’ denote the Encoder output of mixed and reference speech, respectively. ‘\(\textbf{S}_{m}\)’ and ‘\(\textbf{S}_{r}\)’ denote the segmentation result of ‘\(\textbf{E}_{m}\)’ and ‘\(\textbf{E}_{r}\)’, respectively. ‘\(\textbf{D}_{\text{intra}}\)’ and ‘\(\textbf{D}_{\text{inter}}\)’ denote the output of the Intra and Inter Module, respectively.}}
  \label{fig:sefnet}
\end{figure*}
\section{Related Works}
\subsection{Target Speaker Extraction using speaker embeddings}
TSE technology has advanced significantly, with methods generally categorized into time and T-F domain approaches.T-F domain methods typically involve applying the STFT to mixed speech signals and then separating the speech by estimating each speaker’s T-F masks. For instance, early approaches such as VoiceFilter~\cite{wang19h_interspeech,wang2020voicefilter} and SpeakerBeam~\cite{vzmolikova2019speakerbeam} employed deep learning techniques for T-F mask estimation. However, a key challenge with these methods is the phase estimation problem~\cite{luo2018tasnet}.

With the advantages of deep learning in time-domain signal processing, time-domain schemes have gained significant attention. These methods operate directly on the time series, avoiding the phase estimation challenges associated with the STFT. For example, models such as TasNet~\cite{luo2018tasnet} and Conv-TasNet~\cite{luo2019conv} have demonstrated effectiveness in speech separation tasks by using learnable 1-D convolutional neural networks to extract features directly from speech signals. SpEx~\cite{xu2020spex} extended this approach to TSE, improving accuracy and speech quality through multi-scale embedding coefficients and a multi-task learning framework. SpEx+~\cite{ge2020spex+} advanced this further by proposing a comprehensive time-domain speaker extraction solution. By sharing weights between two identical speech encoder networks, SpEx+ addressed the feature space mismatch in SpEx~\cite{xu2020spex}. SpEx$_{pc}$~\cite{wang2021neural} enhanced the utilization of speaker information by integrating speaker embeddings with speech features to predict the target speaker’s mask, exploring how to better leverage speaker information within deep network structures for improved extraction performance. X-SepFormer~\cite{liu2023x} developed an end-to-end TSE model that builds upon X-vectors~\cite{snyder2018x} and SepFormer~\cite{subakan2021attention}. X-SepFormer introduced novel loss schemes that redefine the training objective, focusing on reconstruction performance improvement metrics at the trim segment level. This approach helps the network concentrate on segments with speaker confusion (SC)~\cite{zhao2022target} issues, enhancing system performance while reducing computational costs. X-TF-GridNet~\cite{hao2024x} employs U$^2$-Net~\cite{li2024tabe} and TF-GridNet~\cite{wang2023tf} as the speaker embedding extractor and separator backbone, respectively. Additionally, X-TF-GridNet utilizes adaptive speaker embedding fusion to effectively separate and enhance the target speaker's speech in complex acoustic environments. Despite its successes, this approach may not fully leverage the contextual information available during registration~\cite{yang2024target}.

\subsection{Speaker Embedding Free Target Speaker Extraction}
Traditional TSE methods depend on speaker embeddings, which are fixed-dimensional representations derived from reference speech. Although speaker embeddings capture speaker characteristics effectively, these methods may not fully utilize all the information in the reference speech. Crucial content details, such as local dynamics and temporal structures, are essential for guiding more effective speaker extraction~\cite{hu2024smma,yang2024target}.

Recent advancements have explored embedding-free approaches ~\cite{xiao2019single,yang2023target,zeng2023sef,hu2024smma,yang2024target} that leverage frame-level acoustic features from the reference speech instead of relying on speaker embeddings. This shift addresses the above limitations of traditional methods. These methods do not require a pre-trained speaker embedding extractor or a speaker recognition loss function, focusing instead on directly processing the audio signals to extract the target speaker. SEF-Net~\cite{zeng2023sef} employs a dual-path structure with conformer encoders and a transformer decoder, using cross-multi-head attention to implicitly utilize the speaker information embedded in the reference speech’s conformer encoding outputs. This model achieves performance comparable to other TSE methods without relying on speaker embeddings, offering a novel solution to the embedding mismatch problem. The VE-VE Framework~\cite{yang2023target} designs to handle TSE with ultra-short reference speech. This framework utilizes an RNN-based voice extractor, which relies on RNN states rather than speaker embeddings to capture speaker characteristics. This method addresses the feature fusion problem and effectively supports reference speech, though it is limited to RNN-based extraction networks. With the success of TF-GridNet~\cite{wang2023tf}, there has been a gradual transition from time-domain to T-F domain approaches in TSE. Inspired by SEF-Net~\cite{zeng2023sef}, models such as SMMA-Net~\cite{hu2024smma} and CIENet~\cite{yang2024target} employ attention mechanisms to interact with the T-F representations of both the reference and mixed signals. These models aim to achieve consistent T-F representations that guide more effective extraction by leveraging contextual information directly in the TF domain.

Although these approaches have yielded relatively good results, they have not established a general framework for speaker embedding-free TSE. This paper extends our previous work on speaker embedding-free TSE~\cite{zeng2023sef} to develop a high-performance, universal, and robust framework for TSE.
\begin{figure*}[t!]
  \centering
  \includegraphics[width=0.80\linewidth]{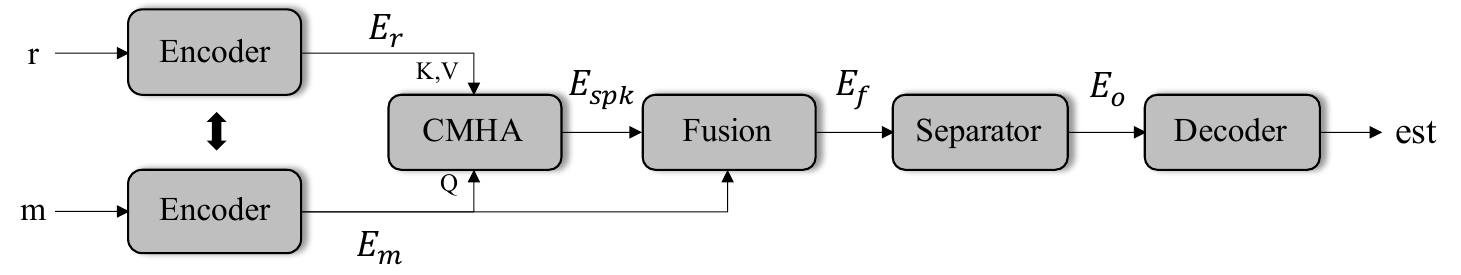}
  \caption{The diagram of the USEF-TSE framework. ‘m’ and ‘r’ denote the mixed speech and \added{reference} speech, respectively. ‘est’ denotes the estimation of the target speaker.}
  \label{fig:usef-tse}
\end{figure*}
\section{Universal Speaker Embedding Free Target Speaker Extraction}
In this section, we will first review our previous work, SEF-Net~\cite{zeng2023sef}. Next, we will introduce the improvements and details of USEF-TSE framework. To validate the universality and effectiveness of USEF-TSE, we will use SepFormer~\cite{subakan2021attention} and TF-GridNet~\cite{wang2023tf} as backbones for the separator. This will involve constructing time-domain and T-F domain TSE models, named USEF-SepFormer and USEF-TFGridNet, respectively.
\subsection{Recap: SEF-Net}
SEF-Net~\cite{zeng2023sef} is a masking-based time-domain TSE network that does not use speaker embeddings. The architecture of SEF-Net~\cite{zeng2023sef} is illustrated in Figure~\ref{fig:sefnet}. SEF-Net employs two weight-sharing convolutional encoders to process the mixed speech \added{\(\textbf{\text{m}}\)} and the reference speech \(\added{\textbf{\text{r}}}\), obtaining STFT-like representations \added{\(\textbf{\text{E}}_{\text{m}} \in \mathbb{R}^{\text{N} \times \text{L}_{1}}\)} and \added{\(\textbf{\text{E}}_{\text{r}} \in \mathbb{R}^{\text{N} \times \text{L}_{2}}\)}. Here, \added{\(\textbf{\text{m}} \in \mathbb{R}^{1 \times \text{T}_{1}}\)} and \added{\(\textbf{\text{r}} \in \mathbb{R}^{1 \times \text{T}_{2}}\)} represent the mixed speech and reference speech, respectively. \added{\text{N}} denotes the feature dimension, while \added{\(\text{L}_{1}\), \(\text{L}_{2}\)} represent the number of frames. In the segmentation stage, \added{\(\textbf{\text{E}}_{\text{m}}\)} and \added{\(\textbf{\text{E}}_{\text{r}}\)} are divided into 3-D features \added{\(\textbf{\text{S}}_{\text{m}} \in \mathbb{R}^{\text{N} \times \text{K} \times \text{S}_{1}}\)} and \added{\(\textbf{\text{S}}_{\text{r}} \in \mathbb{R}^{\text{N} \times \text{K} \times \text{S}_{2}}\)}. Here, K is the length of chunks, and \added{\(\text{S}_{1}\), \(\text{S}_{2}\)} denote the number of chunks. The entire process  of the mixed speech can be formulated as follows:
\begin{equation}
  \added{\textbf{\text{E}}_{\text{m}} = \text{Enc}(\textbf{\text{m}})}
  \label{eq1}
\end{equation}
\begin{equation}
  \added{\textbf{\text{S}}_{\text{m}} = \text{Seg}(\textbf{\text{E}}_{\text{m}})}
  \label{eq2}
\end{equation}
where \added{\(\text{Enc}(\cdot)\)} denotes the convolutional encoder and \added{\(\text{Seg}(\cdot)\)} represents the segmentation operation.

The diagram of the masking network is shown in the center of Figure~\ref{fig:sefnet}. The masking network adopts a dual-path structure similar to SepFormer~\cite{subakan2021attention}, which consists of an intra-module and an inter-module. The details of the intra-module are depicted on the right side of Figure~\ref{fig:sefnet}. This module includes two parallel conformer encoders and a transformer decoder. The weight-shared intra-conformer encoders handle intra-chunk processing for \added{\(\textbf{S}_{\text{m}}\)} and \added{\(\textbf{S}_{\text{r}}\)}. The conformer encoding of \added{\(\textbf{S}_{\text{r}}\)} is then used to query the conformer encoding of \added{\(\textbf{S}_{\text{m}}\)} in the transformer decoder. These conformer encodings are denoted as \added{\(\textbf{C}_{\text{m}}\)} and \added{\(\textbf{C}_{\text{r}}\)}, respectively. In this setup, the cross multi-head attention layer within the transformer decoder functions as a feature fusion module:
\begin{equation}
  \added{\textbf{C}_{\text{m}} = \text{{CE}}(\textbf{S}_{\text{m}}),\textbf{C}_{\text{r}} = \text{{CE}}(\textbf{S}_{\text{r}})}
  \label{eq3}
\end{equation}
%
%
\begin{equation}
  \added{\textbf{D}_{\text{intra}} = \text{TD}(\text{q}=\textbf{C}_{\text{r}}; \text{k},\text{v}=\textbf{C}_{\text{m}})}
  \label{eq4}
\end{equation}
where \added{\(\text{CE}(\cdot)\)} and \added{\(\text{TD}(\cdot)\)} represent the conformer encoder and transformer decoder, respectively. The output of the intra-module is denoted as \added{\(\textbf{D}_{\text{intra}}\)}. The inter-module shares the same components and structure as the intra-module, performing inter-chunk processing on the permuted \added{\(\textbf{D}_{\text{intra}}\)}. The operations in the inter-conformer are the same as the intra-module. \added{\(\textbf{D}_{\text{inter}}\)} denotes the output of the inter-module. After processing through the masking network, SEF-Net~\cite{zeng2023sef} applies an overlap-add operation to \added{\(\textbf{D}_{\text{inter}}\)}, transforming it back into a 2-D feature. Finally, the decoder derives the estimation of the target speaker.

\subsection{Architecture}
Although SEF-Net~\cite{zeng2023sef} achieves performance on par with other TSE methods, it has certain limitations. In SEF-Net~\cite{zeng2023sef}, the transformer decoder is tasked with fusing mixed and reference speech features and extracting the target speaker's speech. This dual responsibility may constrain the model’s separation performance. Furthermore, in Equation~\ref{eq4}, the lengths of \added{\(\textbf{C}_{\text{m}}\)} and \added{\(\textbf{C}_{\text{r}}\)} must be identical, requiring that the mixed speech length \added{\(\text{T}_{1}\)} match the reference speech length \added{\(\text{T}_{2}\)}. This requirement significantly hampers the model’s practical applicability.

To address the aforementioned issues, USEF-TSE improves and extends SEF-Net~\cite{zeng2023sef}. Figure~\ref{fig:usef-tse} illustrates the overall flowchart of USEF-TSE, a universal framework applicable to TSE in both the time and T-F domains. The critical components of USEF-TSE include an Encoder, a CMHA module, a fusion module, a separator, and a decoder.

\subsubsection{Encoder}
USEF-TSE uses the same encoder for both reference and mixed speech. This encoder can be STFT or a one-dimensional convolution, depending on whether the model works in the time or T-F domain.
%
\begin{figure*}[t!]
  \centering
  \includegraphics[width=0.80\linewidth]{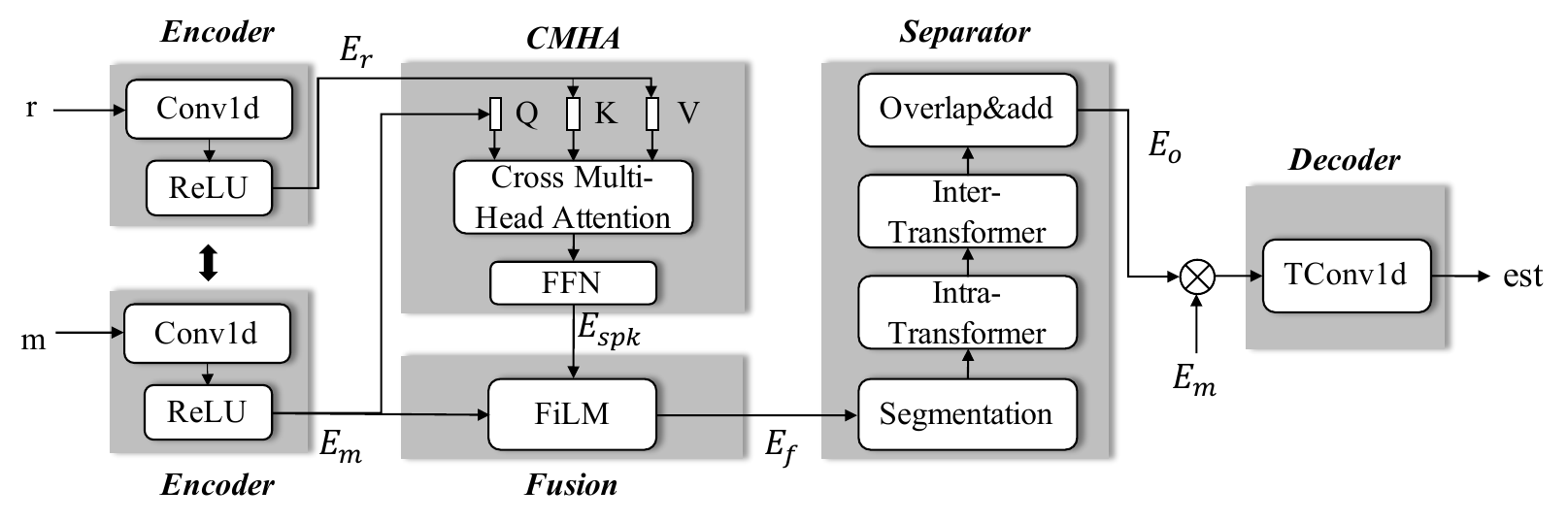}
  \caption{The diagram of USEF-SepFormer network. ‘m’ and ‘r’ denote the mixed speech and \added{reference} speech, respectively. We use two weight sharing encoder to process the mixed and reference speech \added{separately}. ‘est’ denotes the estimation of the target speaker. \(\otimes\) is an operation for element-wise product. The Separator’s parameters are set identically to those of the SepFormer approach.}
  \label{fig:usef-sepformer}
\end{figure*}
\subsubsection{CMHA module}
Unlike SEF-Net~\cite{zeng2023sef}, which combines feature fusion and TSE within the transformer decoder, USEF-TSE separates these tasks to improve performance. The encoder outputs of \added{\(\textbf{m}\)} and \added{\(\textbf{r}\)} are fed into the CMHA module, where a cross multi-head attention mechanism is applied to extract frame-level features of the target speaker:
\begin{equation}
  \added{\textbf{E}_{\text{spk}} = \text{CMHA}(\text{q}=\textbf{E}_{\text{m}}; \text{k},\text{v}=\textbf{E}_{\text{r}})}
  \label{eq5}
\end{equation}
%
where \added{\(\textbf{E}_{\text{m}} \in \mathbb{R}^{\text{N} \times \text{L}_{1}}\)} and \added{\(\textbf{E}_{\text{r}} \in \mathbb{R}^{\text{N} \times \text{L}_{2}}\)} represent the encoder outputs of the mixed speech and reference speech, respectively. The Cross Multi-Head Attention operation is denoted as \added{\(\text{CMHA}(\cdot)\)}, and \added{\(\textbf{E}_{\text{spk}} \in \mathbb{R}^{\text{N} \times \text{L}_{1}}\)} is the output of the CMHA module. Unlike SEF-Net~\cite{zeng2023sef}, the CMHA module in USEF-TSE uses mixed speech encoding as the query. This approach produces a frame-level feature with the same length as \added{\(\textbf{E}_{\text{m}}\)}, allowing the mixed and reference speech lengths to differ in the USEF-TSE framework.

\subsubsection{Fusion module}
USEF-TSE uses \added{\(\textbf{E}_{\text{spk}}\)} as the reference feature for the target speaker and applies a feature fusion module to combine \added{\(\textbf{E}_{\text{spk}}\)} with \added{\(\textbf{E}_{\text{m}}\)}:
\begin{equation}
  \added{\textbf{E}_\text{f} = \text{F}(\textbf{E}_{\text{m}}, \textbf{E}_{\text{spk}})}
  \label{eq6}
\end{equation}
%
where \added{\(\textbf{E}_{\text{f}}\)} denotes the output of the fusion module. The feature fusion, represented by \added{\(\text{F}(\cdot)\)}, can be implemented using methods like feature concatenation or feature-wise linear modulation (FiLM)~\cite{perez2018film}.

\subsubsection{Separator}
\added{\(\textbf{E}_{\text{f}}\)} is fed into the separator to extract the target speaker’s components. The separator can be any deep neural network (DNN) based speech separation backbone, such as DPRNN~\cite{luo2020dual}, SepFormer~\cite{subakan2021attention}, or TF-GridNet~\cite{wang2023tf}. In masking-based approaches, the separator’s output is element-wise multiplied with \added{\(\textbf{E}_{\text{m}}\)} before being passed to the decoder. In mapping-based approaches, the separator’s output is directly sent to the decoder.

\subsubsection{Decoder}
Corresponding to the encoder, the decoder can be either iSTFT or a one-dimensional deconvolution:
\begin{equation}
  \added{\textbf{E}_{\text{o}} = \text{S}(\textbf{E}_{\text{f}})}
  \label{eq7}
\end{equation}
\begin{equation}
  \added{\textbf{est} = \text{Dec}(\text{M}(\textbf{E}_{\text{o}}))}
  \label{eq8}
\end{equation}
where \added{\(\textbf{E}_{\text{o}}\) \(\in\) \(\mathbb{R}^{\text{N} \times \text{L}_{1}}\)} denotes the output of the separator. \added{\(\text{M}(\cdot)\)} refers to the masking or mapping operation. \added{\(\text{S}(\cdot)\)} and \added{\(\text{Dec}(\cdot)\)} represent the separator and decoder, respectively. \added{\(\textbf{est}\) \(\in\) \(\mathbb{R}^{1 \times \text{T}_{1}}\)} denotes the final estimation of the feature from the target speaker.

\subsection{Time and TF Domain USEF-TSE}
Building on the USEF-TSE framework, we propose two TSE models: USEF-SepFormer and USEF-TFGridNet. These models employ the SepFormer~\cite{subakan2021attention} and TF-GridNet~\cite{wang2023tf} backbone networks as separators in the USEF-TSE architecture . Accordingly, USEF-SepFormer operates in the time domain, while USEF-TFGridNet operates in the T-F domain for TSE.
\subsubsection{USEF-SepFormer}
The diagram of the USEF-SepFormer is shown in the Figure~\ref{fig:usef-sepformer}. In this model,  two weight sharing encoders are used to process the mixed and reference speech:
\begin{equation}
  \added{\textbf{E}_{\text{m}} = \text{ReLU}(\text{Conv1d}(\textbf{m}))}
  \label{eq9}
\end{equation}
\begin{equation}
  \added{\textbf{E}_{\text{r}} = \text{ReLU}(\text{Conv1d}(\textbf{f}))}
  \label{eq10}
\end{equation}
where \added{\(\textbf{E}_{\text{m}}\) \(\in\) \(\mathbb{R}^{\text{B} \times \text{N} \times \text{L}_{1}}\)} and \added{\(\textbf{E}_{\text{r}}\) \(\in\) \(\mathbb{R}^{\text{B} \times \text{N} \times \text{L}_{2}}\)} represent the encoded outputs of the mixed and reference speech, respectively. B is the batch size. N is the feature dimension. \added{\(\text{L}_{1}\)} and \added{\(\text{L}_{2}\)} are the number of time steps. Using the CMHA module as described in Equation~\ref{eq5}, \added{\(\textbf{E}_{\text{m}}\)} and \added{\(\textbf{E}_{\text{r}}\)} are processed to produce \added{\(\textbf{E}_{\text{spk}}\) \(\in\) \(\mathbb{R}^{\text{B} \times \text{N} \times \text{L}_{1}}\)}. To align with the SepFormer~\cite{subakan2021attention} backbone network, we employ a transformer encoder as the CMHA module in USEF-SepFormer. Subsequently, \added{\(\textbf{E}_{\text{spk}}\)} is combined with \added{\(\textbf{E}_{\text{m}}\)} through the fusion module. In USEF-SepFormer, FiLM is utilized for feature fusion:
\begin{equation}
  \added{\textbf{E}_{\text{f}} = \text{FiLM}(\textbf{E}_{\text{m}}, \textbf{E}_{\text{spk}})}
  \label{eq11}
\end{equation}
\begin{equation}
  \added{\text{FiLM}(\textbf{E}_{\text{m}}, \textbf{E}_{\text{spk}}) = \gamma(\textbf{E}_{\text{spk}}) \cdot \textbf{E}_{\text{m}} + \beta(\textbf{E}_{\text{spk}})}
  \label{eq12}
\end{equation}
where \added{\(\textbf{E}_{\text{f}}\) \(\in\) \(\mathbb{R}^{\text{B} \times \text{N} \times \text{L}_{1}}\)} represents the fused features. The scaling and shifting vectors of FiLM are represented by \(\gamma(\cdot)\) and \(\beta(\cdot)\), respectively. \added{\(\textbf{E}_{\text{f}}\)} is then input into the separator.

USEF-SepFormer utilizes the SepFormer~\cite{subakan2021attention} backbone network as its separator. SepFormer~\cite{subakan2021attention} is is a masking-based, dual-path structure time-domain speech separation model. It divides long sequence features into equal-length chunks and performs intra-chunk and inter-chunk operations on these segmented features:
\begin{equation}
  \added{\textbf{E}_{\text{intra}} = \textbf{S}_{\text{f}} + \text{IntraTE}(\textbf{S}^{\prime}_{\text{f}})}
  \label{eq13}
\end{equation}
\begin{equation}
  \added{\textbf{E}_{\text{inter}} = \textbf{E}_{\text{intra}} + \text{InterTE}(\textbf{E}^{\prime}_{\text{intra}})}
  \label{eq14}
\end{equation}
where \added{\(\textbf{S}_{\text{f}}\) \(\in\) \(\mathbb{R}^{\text{B} \times \text{N} \times \text{K} \times \text{S}}\)} denotes the segmentation result of \added{\(\textbf{E}_{\text{f}}\)}. \added{\(\textbf{S}^{\prime}_{\text{f}}\) \(\in\) \(\mathbb{R}^{(\text{B} \times \text{S}) \times \text{K} \times \text{N}}\)} is a transformation of \added{\(\textbf{S}_{\text{f}}\)}. \added{\(\textbf{E}_{\text{intra}}\), \(\textbf{E}_{\text{inter}}\) \(\in\) \(\mathbb{R}^{\text{B} \times \text{N} \times \text{K} \times \text{S}}\)} are the outputs of the intra-chunk and inter-chunk modules, respectively. \added{\(\textbf{E}^{\prime}_{\text{intra}}\) \(\in\) \(\mathbb{R}^{(\text{B} \times \text{K}) \times \text{S} \times \text{N}}\)} is a transformation of \added{\(\textbf{E}_{\text{intra}}\)}. \added{\(\text{IntraTE}(\cdot)\)} and \added{\(\text{InterTE}(\cdot)\)} are the intra and inter transformer encoder, respectively. After the dual-path network, USEF-SepFormer perform an overlap-add operation on \added{\(\textbf{E}_{\text{inter}}\)} to obtain \added{\(\textbf{E}_{\text{o}}\) \(\in\) \(\mathbb{R}^{\text{B} \times \text{N} \times \text{L}_{1}}\)}. \added{\(\textbf{E}_{\text{o}}\)} is the estimated mask for the target speaker. At last, the decoder processes the product of \added{\(\textbf{E}_{\text{m}}\)} and \added{\(\textbf{E}_{\text{o}}\)} and yielding the estimated target speaker speech:
\begin{equation}
  \added{\textbf{est} = \text{TConv1d}(\textbf{E}_{\text{m}} * \textbf{E}_{\text{o}})}
  \label{eq15}
\end{equation}
where \added{\(\text{TConv1d}(\cdot)\)} denotes the one-dimensional transposed convolution operation. \added{\(\textbf{est}\) \(\in\) \(\mathbb{R}^{\text{B} \times 1 \times \text{T}_{1}}\)} denotes the estimation of the speech from the target speaker.
\begin{figure*}[t!]
  \centering
  \includegraphics[width=0.80\linewidth]{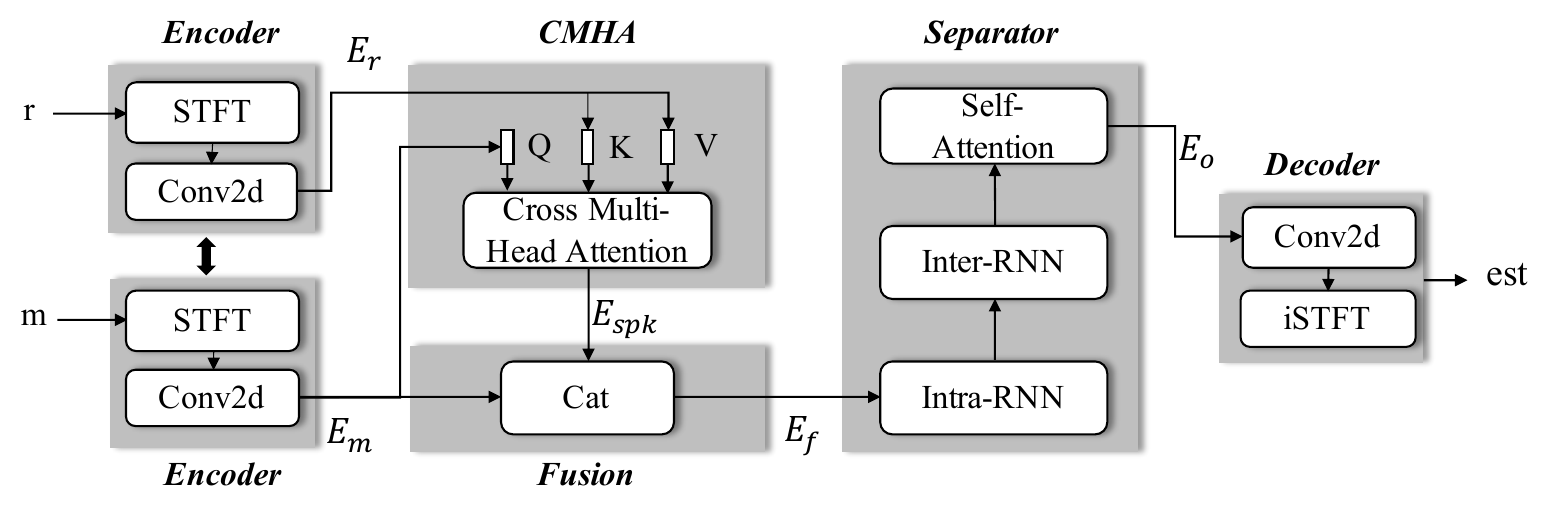}
  \caption{The diagram of USEF-TFGridNet network. We use the same STFT settings to extract the acoustic features of both the mixed and the reference speech. The weight shared Conv2d increase the number of channel for both the mixed and reference speech' acoustic features. ‘m’ and ‘r’ denote the mixed speech and \added{reference} speech, respectively. ‘est’ denotes the estimation of the target speaker. The Separator’s parameters are set identically to those of the TF-GridNet approach.}
  \label{fig:usef-tfgridnet}
\end{figure*}
\subsubsection{USEF-TFGridNet}
The diagram of USEF-TFGridNet is shown in Figure~\ref{fig:usef-tfgridnet}. USEF-TFGridNet uses two weight-sharing 2D convolutions to process the real and imaginary (RI) components of the STFT features for both the mixed speech and the reference speech:
\begin{equation}
  \added{\textbf{E}_{\text{m}} = \text{Conv2d}(\textbf{RI}_{\text{m}})}
  \label{eq16}
\end{equation}
\begin{equation}
  \added{\textbf{E}_{\text{r}} = \text{Conv2d}(\textbf{RI}_{\text{r}})}
  \label{eq17}
\end{equation}
where \added{\(\text{Conv2d}(\cdot)\)} refers to 2D convolutions. \added{\(\textbf{RI}_{\text{m}}\) \(\in\) \(\mathbb{R}^{\text{B} \times 2 \times \text{F} \times \text{L}_{1}}\)} and \added{\(\textbf{RI}_{\text{r}}\) \(\in\) \(\mathbb{R}^{\text{B} \times 2 \times \text{F} \times \text{L}_{2}}\)} represent the stacked real and imaginary components of the STFT features for the mixed and reference speech, respectively, respectively. F is the embedding dimension for each T-F unit. \added{\(\text{L}_{1}\)} and \added{\(\text{L}_{2}\)} are the numbers of T-F units. The encoder results for the mixed and reference speech are \added{\(\textbf{E}_{\text{m}}\) \(\in\) \(\mathbb{R}^{\text{B} \times \text{C} \times \text{F} \times \text{L}_{1}}\)} and \added{\(\textbf{E}_{\text{r}}\) \(\in\) \(\mathbb{R}^{\text{B} \times \text{C} \times \text{F} \times \text{L}_{2}}\)}. Through Equation (\ref{eq5}), the CMHA module processes \added{\(\textbf{E}_{\text{m}}\)} and \added{\(\textbf{E}_{\text{r}}\)}, outputting \added{\(\textbf{E}_{\text{spk}}\) \(\in\) \(\mathbb{R}^{\text{B} \times \text{C} \times \text{F} \times \text{L}_{1}}\)}. To align with the separator, USEF-TFGridNet uses the attention module from TF-GridNet~\cite{wang2023tf} as its CMHA module. Next, we use concatenation as the feature fusion method to combine \added{\(\textbf{E}_{\text{m}}\)} and \added{\(\textbf{E}_{\text{spk}}\)}, resulting in the fused feature \added{\(\textbf{E}_{\text{f}}\) \(\in\) \(\mathbb{R}^{\text{B} \times \text{2C} \times \text{F} \times \text{L}_{1}}\)}. 

USEF-TFGridNet employs the TF-GridNet~\cite{wang2023tf} backbone network as its separator. TF-GridNet~\cite{wang2023tf} is a mapping-based T-F domain speech separation model. The backbone network of TF-GridNet~\cite{wang2023tf} is composed of three key modules: the intra-frame full-band module, the sub-band temporal module, and the cross-frame self-attention module. The intra-frame full-band module processes data along the F dimension to capture full-band spectral and spatial information:
\begin{equation}
  \added{\textbf{U}_{\text{f}} = \text{Unfold}(\text{V}(\textbf{E}^\prime_{\text{f}}))}
  \label{eq18}
\end{equation}
\begin{equation}
  \added{\textbf{E}_{\text{intra}} = \textbf{E}^\prime_{\text{f}} + \text{V}(\text{TConv1d}(\text{BLSTM}(\textbf{U}_{\text{f}})))}
  \label{eq19}
\end{equation}
where \added{\(\textbf{E}^\prime_{\text{f}}\) \(\in\) \(\mathbb{R}^{\text{B} \times \text{2C} \times \text{F}^\prime \times \text{L}_{1}^\prime}\)} represents the zero-padded version of \added{\(\textbf{E}_{\text{f}}\)}. The reshaping operation, denoted as \added{\(\text{V}(\cdot)\)}, reshapes \added{\(\textbf{E}^\prime_{\text{f}}\)} into the shape \added{\(\mathbb{R}^{(\text{B} \times \text{L}_{1}^\prime) \times \text{F}^\prime \times \text{2C}}\)}. Here, \added{\(\text{F}^\prime = \lceil\frac{\text{F}-\text{ks}}{\text{hs}}\rceil \times \text{hs} + \text{ks}\)}. The function \added{\(\text{Unfold}(\cdot)\)} corresponds to the torch.unfold function, with \added{\(\textbf{U}_{\text{f}}\) \(\in\) \(\mathbb{R}^{(\text{B} \times \text{L}_{1}^\prime) \times (\frac{\text{F}^\prime-\text{ks}}{\text{hs}}+1) \times (\text{2C} \times \text{ks})}\)} as its output. The parameters \added{\(\text{ks}\), \(\text{hs}\)} are the kernel size and stride of torch.unfold function, respectively. \added{\(\text{TConv1d}(\cdot)\)} denotes a one-dimensional transposed convolution with kernel size \added{\(\text{ks}\)} and stride \added{\(\text{hs}\)}. \added{\(\text{BLSTM}(\cdot)\)} means the bidirectional long short-term memory (BLSTM)~\cite{hochreiter1997long} layer. The output of the intra-frame full-band module is \added{\(\textbf{E}_{\text{intra}}\) \(\in\) \(\mathbb{R}^{\text{B} \times \text{2C} \times \text{F}^\prime \times \text{L}_{1}^\prime}\)}. The sub-band temporal module then processes \added{\(\textbf{E}_{\text{intra}}\)} along the \added{\(\text{L}_{1}^\prime\)} dimension in the same manner as the intra-frame full-band module:
\begin{equation}
  \added{\textbf{E}_{\text{sub}} = \textbf{E}_{\text{intra}} + \text{V}(\text{Sub}(\textbf{E}^\prime_{\text{intra}}))}
  \label{eq20}
\end{equation}
where \added{\(\textbf{E}^\prime_{\text{intra}}\) \(\in\) \(\mathbb{R}^{(\text{B} \times \text{L}_{1}^\prime) \times \text{F}^\prime \times \text{2C}}\)} represents a transformation of \added{\(\textbf{E}_{\text{intra}}\)}. The function \added{\(\text{Sub}(\cdot)\)} denotes the operations within the sub-band temporal module, following the same procedures as in Equations (\ref{eq18}) - (\ref{eq19}). The output of the sub-band temporal module is  \added{\(\textbf{E}_{\text{sub}}\) \(\in\) \(\mathbb{R}^{\text{B} \times \text{2C} \times \text{F}^\prime \times \text{L}_{1}^\prime}\)}. After removing zero padding, \added{\(\textbf{E}_{\text{sub}}\)} is passed into the cross-frame self-attention module. This module is the same as the one in TF-GridNet~\cite{wang2023tf}. It uses multi-head attention, allowing each T-F unit to attend directly to other relevant units:
\begin{equation}
  \added{\textbf{E}_{\text{cross}} = \text{Cross}(\textbf{E}_{\text{sub}}[:,:,\text{F},\text{L}_{1}])}
  \label{eq21}
\end{equation}
\begin{equation}
  \added{\textbf{E}_{\text{o}} = \textbf{E}_{\text{cross}} + \textbf{E}_{\text{sub}}[:,:,\text{F},\text{L}_{1}]}
  \label{eq22}
\end{equation}
where \added{\(\text{Cross}(\cdot)\)} refers to the operations in the cross-frame self-attention module, which are the same as in TF-GridNet~\cite{wang2023tf}. \added{\(\textbf{E}_{\text{Cross}}\), \(\textbf{E}_{\text{o}}\) \(\in\) \(\mathbb{R}^{\text{B} \times \text{2C} \times \text{F} \times \text{L}_{1}}\)} are the outputs of the cross-frame self-attention module and the separator, respectively. A 2D transposed convolution layer is then applied to adjust the number of channels in \added{\(\textbf{E}_{\text{o}}\)} back to 2. Finally, the result of the 2D transposed convolution is converted back to a waveform using iSTFT.
\begin{equation}
  \added{\textbf{est} = \text{iSTFT}(\text{TConv2d}(\textbf{E}_{\text{o}}))}
  \label{eq23}
\end{equation}
where \added{\(\text{TConv2d}(\cdot)\)} denotes the 2D transposed convolution operation. \added{\(\textbf{est}\) \(\in\) \(\mathbb{R}^{\text{B} \times 1 \times \text{T}_{1}}\)} denotes the estimation of the target speaker.
\begin{table}[]
\caption{Hyperparameters of the USEF-SepFormer.}
  \label{tab:hyp-usef-sepformer}
  \centering
\begin{tabular}{c|c}
\hline
\textbf{Symbol} & \textbf{Description}                                                                                                                       \\ \hline
\added{C$_l$}     & The number of the attention layers in the CMHA module                                                                             \\ \hline
C$_h$    & \begin{tabular}[c]{@{}c@{}}The number of the attention heads \\ in the CMHA module\end{tabular}                                   \\ \hline
C$_d$     & \begin{tabular}[c]{@{}c@{}}The dimension of the feed-forward network \\ in the CMHA module\end{tabular}                           \\ \hline
S$_l$    & \begin{tabular}[c]{@{}c@{}}The number of the attention layers \\ in the intra- and inter- transformer encoder\end{tabular}        \\ \hline
S$_h$     & \begin{tabular}[c]{@{}c@{}}The number of the attention heads \\ in the intra- and inter- transformer encoder\end{tabular}         \\ \hline
S$_d$     & \begin{tabular}[c]{@{}c@{}}The dimension of the feed-forward network \\ in the intra- and inter- transformer encoder\end{tabular} \\ \hline
\end{tabular}
\end{table}
\section{Experimental Setup}
\subsection{Datasets}
\subsubsection{Anechoic Speech Mixtures}
We validate our proposed methods on the WSJ0-2mix dataset~\cite{hershey2016deep}, a benchmark for speech separation and TSE in anechoic conditions. WSJ0-2mix is a two-speaker mixed dataset derived from the Wall Street Journal (WSJ0) corpus~\cite{garofolo1993csr}. It consists of three subsets: the training set with 20,000 utterances from 101 speakers, the development set with 5,000 utterances from 101 speakers, and the test set with 3,000 utterances from 18 speakers. The training and development sets feature speakers from WSJ0 “si\_tr\_s”, while the test set includes speakers from WSJ0 “si\_dt\_05” and “si\_et\_05”. The speakers in the training and development sets are different from those in the test set. Utterances from two speakers are randomly selected to create mixed audio with a relative signal-to-noise ratio (SNR) between 0 dB and 5 dB. During training, the reference speech for the target speaker is randomly chosen and updated in each epoch. In the inference stage, we use the same reference speech as SpEx+~\footnotemark[2]\footnotetext[2]{\url{https://github.com/gemengtju/SpEx_Plus}}.

\subsubsection{Noisy Speech Mixtures}
We use the WSJ0 Hipster Ambient Mixtures (WHAM!) dataset~\cite{Wichern2019WHAM} to evaluate our methods for noisy TSE. WHAM! is a noisy variant of the WSJ0-2mix dataset, featuring background noise recorded in urban environments such as coffee shops and parks. It preserves the same relative levels between the two speakers as WSJ0-2mix. Noise is added by sampling a random SNR value from a uniform distribution ranging from -6 to +3 dB. During training, the reference speech for the target speaker is randomly selected and updated each epoch. In the inference stage, we use the same reference speech as in the anechoic scenario.

\subsubsection{Noisy-Reverberant Speech Mixtures}
We utilize the WHAMR! dataset~\cite{Maciejewski2020WHAMR} to test our algorithms for noisy reverberant TSE. WHAMR! extends the WSJ0-2mix dataset by reverberating each clean source and incorporating background noise. The reverberation time (T60) in these mixtures ranges from 0.2 to 1.0 seconds. The signal-to-noise ratio (SNR) between the louder speaker and the noise varies from -6 to 3 dB, the relative energy level between the two speakers ranges from -5 to 5 dB, and the speaker-to-array distance spans from 0.66 to 2.0 meters. The dataset includes 20,000 mixtures for training, 5,000 mixtures for development, and 3,000 mixtures for testing. During the training, the direct-path signal from each speaker, recorded by the first microphone, serves as the target. This signal is also used as the reference for metric evaluation. The target speaker’s reference speech is randomly selected and updated each epoch during the training. We use the same reference speech as in the anechoic scenario for inference.
\begin{table}[]
\caption{Hyperparameters of the USEF-TFGridNet.}
  \label{tab:hyp-usef-tfgridnet}
  \centering
\begin{tabular}{c|c}
\hline
\textbf{Symbol} & \textbf{Description}                                                                                                                          \\ \hline
C$_l$      & The number of the attention layers in the CMHA module                                                                               \\ \hline
C$_h$     & \begin{tabular}[c]{@{}c@{}}Number of the attention heads \\ in the CMHA module\end{tabular}                                         \\ \hline
C$_d$     & \begin{tabular}[c]{@{}c@{}}Dimension of the feed-forward network \\ in the CMHA module\end{tabular}                                 \\ \hline
K      & Kernel size for Unfold and TConv1d                                                                                                  \\ \hline
J      & Stride size for Unfold and TConv1d                                                                                                  \\ \hline
H      & \begin{tabular}[c]{@{}c@{}}Number of hidden units of BLSTMs \\ in the TF-GridNet blocks\end{tabular}                                \\ \hline
E      & \begin{tabular}[c]{@{}c@{}}Number of output channels in point-wise Conv2D\\ in the TF-GridNet blocks' attention module\end{tabular} \\ \hline
S$_l$     & \begin{tabular}[c]{@{}c@{}}Number of the attention layers \\ in the TF-GridNet blocks' attention module\end{tabular}                \\ \hline
S$_h$     & \begin{tabular}[c]{@{}c@{}}Number of the attention heads \\ in the TF-GridNet blocks' attention module\end{tabular}                 \\ \hline
S$_d$     & \begin{tabular}[c]{@{}c@{}}Dimension of the feed-forward network \\ in the TF-GridNet blocks' attention module\end{tabular}         \\ \hline
\end{tabular}
\end{table}
\subsubsection{\added{LibriMix}}
\added{LibriMix~\cite{cosentino2020LibriMix} is used to evaluate the stability of USEF-TSE on the more diverse datasets. The LibriMix dataset, derived from LibriSpeech~\cite{7178964} and WHAM! noise~\cite{Wichern2019WHAM}, is one of the benchmark resources for SS and TSE tasks. It comprises two main subsets: Libri2Mix and Libri3Mix. The dataset is available in four variations, defined by two sampling rates (16 kHz and 8 kHz) and two mixing modes (min and max). In the min mode, the mixture is truncated to the duration of the shortest utterance, whereas in the max mode, the shortest utterance is zero-padded to match the length of the longest one. In our experimental setup, we employ the 100-hour training subsets from Libri2Mix and Libri3Mix, comprising 98 hours of mixtures from 251 unique speakers after pre-processing. Evaluation is conducted on their respective test sets containing 11 hours of data from 40 speakers. We exclusively adopt the max mode version at a 16 kHz sampling rate for both training and inference stages.}

\subsubsection{\added{ICASSP 2023 DNS Challenge Dataset}}
\added{The blind test dataset from the ICASSP 2023 DNS Challenge~\cite{10474162} is used to evaluate the generalizability of our proposed USEF-TSE framework on out-of-domain datasets. The blind test set includes audio recordings from two different conditions: headset and speakerphone. For the Headset track, the dataset contains 389 real test clips, of which 220 have interfering talkers and 51 are leakage clips. Similarly, the Speakerphone track's dataset consists of 331 real test clips, of which 220 have interfering speakers and 51 are leakage clips. In our experiments, we randomly extract a 15-second segment from the provided enrollment speech as the reference audio. Both the enrollment speech and the mixed speech are resampled to 8 kHz.}

\subsection{Network Configuration}
\subsubsection{USEF-SepFormer}
We configure USEF-SepFormer based on the parameter settings of SepFormer from SpeechBrain~\cite{speechbrain}. The encoder employs a 1D convolutional layer with a kernel size of 16 and a stride of 8. It has an input dimension of 1 and an output dimension of 256. The transformer encoder in the CMHA module consists of 4 layers, 8 parallel attention heads, and a 1024-dimensional feed-forward network. FiLM is implemented with two single-layer linear layers for scaling and shifting, with an input and output dimension of 256. The segmentation stage divides the input into chunks of size K = 250. For the SepFormer backbone network, the intra-transformer encoder and inter-transformer encoder each consist of 8 layers, 8 parallel attention heads, and a 1024-dimensional feed-forward network. The dual-path processing pipeline is repeated twice. The decoder uses a 1D transposed convolutional layer with the same kernel size and stride as the encoder. The hyperparameters for USEF-SepFormer are summarized in Table~\ref{tab:hyp-usef-sepformer}.
\begin{table*}
\caption{\added{SDRi(dB), SI-SDRi(dB) and computational cost of the USEF-TSE method compared with speaker embedding based methods on the WSJ0-2mix dataset. We report the number of model parameters in millions (M).}}
  \label{tab:cmp-cost}
  \centering
\begin{tabular}{c|c|ccccc}
\hline
\textbf{\added{Model}}    & \textbf{\begin{tabular}[c]{@{}c@{}}\added{Using Speaker}\\ \added{Embedding?}\end{tabular}} & \textbf{\added{SDRi(dB)}} & \textbf{\added{SI-SDRi(dB)}} & \textbf{\added{Parameter(M)}} & \textbf{\added{GMAC/S}} & \textbf{\textbf{\begin{tabular}[c]{@{}c@{}}\added{Traing Time}\\ \added{(min/epoch)}\end{tabular}}} \\ \hline
\added{Res-SepFormer-v1} & \added{Yes}                                                                           & \added{17.3}              & \added{16.5}                 & \added{17.6}                    & \added{43.3}                & \added{125.5}                       \\
\added{Res-SepFormer-v2} & \added{Yes}                                                                           & \added{17.9}              & \added{17.1}                 & \added{19.7}                    & \added{45.4}                & \added{128.6}                       \\
\added{X-SepFormer\cite{liu2023x}}       & \added{Yes}                                                                           & \added{19.7}               & \added{19.0}                   & \added{$17.9^\ddag$}                   & \added{$43.9^\ddag$}               & \added{$127.3^\ddag$}                      \\
\added{USEF-SepFormer}    & \added{No}                                                                          & \added{20.3}               & \added{19.9}                  & \added{19.7}                & \added{45.8}            & \added{147.9}                       \\
\added{Res-TFGridNet-v1} & \added{Yes}                                                                           & \added{20.1}              & \added{19.5}                 &  \added{10.6}                   & \added{86.7}                &  \added{111.4}                      \\
\added{Res-TFGridNet-v2} & \added{Yes}                                                                           & \added{20.6}              & \added{19.9}                 & \added{15.2}                    & \added{124.5}                & \added{163.4}                       \\
\added{X-TF-GridNet\cite{hao2024x}}      & \added{Yes}                                                                           & \added{21.7}               & \added{20.7}                  & \added{12.7}                & \added{113.2}               & \added{$366.4^\ddag$}                      \\
\added{USEF-TFGridNet}    & \added{No}                                                                          & \added{23.5}               & \added{23.3}                  & \added{15.2}                & \added{125.3}           & \added{178.8}                        \\ \hline
\end{tabular}
\begin{minipage}{0.9\textwidth}
        \footnotesize \textit{\added{Notes:} \\
        \added{(a) GMAC/s is computed based on a 1-second mixed speech segment and 8.6-second reference speech.} \\
        \added{(b) For the training time, we report the duration required to complete one epoch using 40,000 4-second segments on an NVIDIA RTX A6000 GPU with 48 GB of memory. The batch size is adjusted to maximize GPU memory utilization.} \\
        \added{(c) “$\ddag$” indicates that the results come from our own implemented experiments. Specifically, X-SepFormer is reimplemented through~\cite{liu2023x}, while X-TF-GridNet is based on the code available at: \url{https://github.com/HaoFengyuan/X-TF-GridNet}}
        }
    \end{minipage}
\end{table*}
\subsubsection{USEF-TFGridNet}
We configure USEF-TFGridNet based on the parameter settings of TF-GridNet from ESPNet~\cite{watanabe2018espnet}. For STFT, the window length is 16 ms, and the hop length is 8 ms. A 128-point Fourier transform is used to extract 65-dimensional complex STFT features at each frame. The 2D convolution layer has a kernel size of (3,3) and a stride of 1, with input and output dimensions of 2 and 128, respectively. The attention block in the CMHA module consists of 1 layer, 4 parallel attention heads, and a 512-dimensional feed-forward network. The kernel size and stride of the torch.unfold function are both 1. In the intra-frame full-band and sub-band temporal modules, the BLSTM layer has 256 units. The cross-frame self-attention module uses 1 layer, 4 parallel attention heads, and a 512-dimensional feed-forward network. The number of TF-GridNet blocks is 6. The 2D transposed convolution layer has the same kernel size and stride as the 2D convolution layer, with input and output dimensions of 256 and 2, respectively. The hyperparameters for USEF-TFGridNet are summarized in Table~\ref{tab:hyp-usef-tfgridnet}.

\subsubsection{\added{Speaker Embedding based SepFormer and TF-GridNet}}
\added{We employ ResNet34~\cite{10446780} as the speaker embedding extractor to develop four target speaker extraction models: Res-SepFormer variants (v1/v2) and Res-TFGridNet variants (v1/v2), which integrate SepFormer and TF-GridNet, respectively. Res-SepFormer-v1 and Res-TFGridNet-v1 adopt the conventional target speaker extraction framework (as shown in Figure~\ref{fig:tytse}). In contrast, Res-SepFormer-v2 and Res-TFGridNet-v2 utilize the same backbone as USEF-SepFormer and USEF-TFGridNet, respectively (as shown in Figure~\ref{fig:usef-sepformer} and Figure~\ref{fig:usef-tfgridnet}). It is important to note that Res-SepFormer-v1 and Res-TFGridNet-v1 employ a simple concatenation-based feature fusion approach. As a result, Res-SepFormer-v1 and Res-TFGridNet-v1 have a smaller model size and lower computational complexity than Res-SepFormer-v2 and Res-TFGridNet-v2.}

\subsection{Training Details}
We trained our models using the Adam optimizer~\cite{KingBa15}, with an initial learning rate set to 1e-4. The learning rate was halved if the validation loss did not improve within 3 epochs. No dynamic mixing or data augmentation was applied during training. The speech clips were truncated to 4 seconds during training, while the full speech clips were evaluated during inference. In the evaluation phase, each speaker in the mixed speech is considered as the target speaker in turn. The models were trained to maximize the scale-invariant signal-to-distortion ratio (SI-SDR)~\cite{le2019sdr}, which is defined as follows:
\begin{equation}
\begin{cases}
 \added{\textbf{s}_{\text{T}} = \frac{<\hat{\textbf{s}},\textbf{s}>\textbf{s}}{||\textbf{s}||^2}}\\ 
\added{\textbf{s}_{\text{E}} = \hat{\textbf{s}} - \textbf{s}_{\text{T}}} \\
\added{\text{SI-SDR} = -10\lg{\frac{||\textbf{s}_{\text{T}}||^2}{||\textbf{s}_{\text{E}}||^2}}}
\end{cases}
\label{eq24}
\end{equation}
where \added{\(\hat{\textbf{s}}\) \(\in\) \(\mathbb{R}^{1 \times \text{T}}\)} represents the estimated target speaker speech, while \added{\(\textbf{s}\) \(\in\) \(\mathbb{R}^{1 \times \text{T}}\)} represents the clean source speech. \added{\(<\textbf{s},\textbf{s}>\)} denotes the power of the signal \added{\(\textbf{s}\)}.

\subsection{Evaluation Metrics}
We use SI-SDR or SI-SDR improvement (SI-SDRi)~\cite{le2019sdr} and signal-to-distortion ratio (SDR) or SDR improvement (SDRi)~\cite{vincent2006performance} as objective metrics to assess the accuracy of TSE. In addition to distortion metrics, we evaluate the estimated target speaker’s speech quality using the perceptual evaluation of subjective quality (PESQ)~\cite{rix2001perceptual}. The PESQ values are calculated using the pypesq\footnotemark[3]\footnotetext[3]{\url{https://github.com/youngjamespark/python-pypesq}} tookit, and the thop\footnotemark[4]\footnotetext[4]{\url{https://github.com/Lyken17/pytorch-OpCounter}} toolkit is employed to measure the number of model parameters and computational cost. \added{We use DNSMOS~\cite{reddy2022dnsmos} to evaluate the performance of TSE models on the blind test set from the ICASSP 2023 DNS Challenge.}

%
\begin{table*}
\caption{SDRi(dB), SI-SDRi(dB) and PESQ of of the USEF-SepFormer with Different Hyper-Parameters. We report the number of model parameters in millions (M).}
  \label{tab:abl-usef-sepformer}
  \centering
\begin{tabular}{c|c|c|cccccc|c|c|c|c}
\hline
\textbf{Row} & \textbf{System}         & \begin{tabular}[c]{@{}c@{}}\textbf{Use Speaker}\\ \textbf{Embedding?}\end{tabular} & \added{\textbf{C$_l$}} & \added{\textbf{C$_h$}} & \added{\textbf{C$_d$}}   & \added{\textbf{S$_l$}} & \added{\textbf{S$_h$}} & \added{\textbf{S$_d$}}   & \textbf{SDRi(dB)} & \textbf{SI-SDRi(dB)} & \textbf{PESQ} & \textbf{Parameter(M)} \\ \hline
1   & Mixture        & -                                                                & -  & -  & -    & -  & -  & -    & -         & -            & 2.01 & -            \\
2   & X-SepFormer\cite{liu2023x}    & Yes                                                              & -  & -  & -    & -  & -  & - & 19.7      & 19.1         & -    & -            \\
3   & USEF-SepFormer & No                                                               & 1  & 4  & 1024 & 4  & 8  & 2048 & 18.2      & 18.0         & 3.50 & 18.2         \\
4   & USEF-SepFormer & No                                                               & 1  & 4  & 1024 & 8  & 8  & 1024 & 18.1      & 17.8         & 3.48 & 18.2         \\
5   & USEF-SepFormer & No                                                               & 1  & 8  & 1024 & 4  & 8  & 2048 & 18.5      & 18.2         & 3.53 & 18.2         \\
6   & USEF-SepFormer & No                                                               & 1  & 8  & 1024 & 8  & 8  & 1024 & 19.3      & 19.0         & 3.60 & 18.2         \\
7   & USEF-SepFormer & No                                                               & 4  & 4  & 1024 & 4  & 8  & 2048 & 19.1      & 18.8         & 3.59 & 19.7         \\
8   & USEF-SepFormer & No                                                               & 4  & 4  & 1024 & 8  & 8  & 1024 & 19.5      & 19.2         & 3.61 & 19.7         \\
9   & USEF-SepFormer & No                                                               & 4  & 8  & 1024 & 4  & 8  & 2048 & 18.7      & 18.4         & 3.55 & 19.7         \\
10   & USEF-SepFormer & No                                                               & 4  & 8  & 1024 & 8  & 8  & 1024 & 20.3      & 19.9         & 3.67 & 19.7         \\ \hline
\end{tabular}
\end{table*}
\begin{table*}
\caption{SDRi(dB), SI-SDRi(dB) and PESQ of of the USEF-TFGridNet with Different Hyper-Parameters. We report the number of model parameters in millions (M).}
  \label{tab:abl-usef-tfgridnet}
  \centering
\begin{tabular}{c|c|c|cccccc|c|c|c|c}
\hline
\textbf{Row} & \textbf{System}         & \begin{tabular}[c]{@{}c@{}}\textbf{Use Speaker}\\ \textbf{Embedding?}\end{tabular} & \added{\textbf{C$_l$}} & \added{\textbf{C$_h$}} & \added{\textbf{C$_d$}}   & \textbf{K} & \textbf{J} & \textbf{E}   & \textbf{SDRi(dB)} & \textbf{SI-SDRi(dB)} & \textbf{PESQ} & \textbf{Parameter(M)} \\ \hline
1   & Mixture        & -                                                                & -  & -  & -    & - & - & -   & -         & -            & 2.01 & -            \\
2   & X-TF-GridNet\cite{hao2024x}   & Yes                                                              & -  & -  & -    & - & - & -  & 21.7      & 20.7         & 3.70 & 20.7         \\
3   & USEF-TFGridNet & No                                                               & 1  & 4  & 512  & 1 & 1 & 128 & 23.5      & 23.3         & 3.92 & 15.2         \\
4   & USEF-TFGridNet & No                                                               & 1  & 4  & 512  & 4 & 1 & 32  & 22.7      & 22.5         & 3.88 & 14.3         \\
5   & USEF-TFGridNet & No                                                               & 1  & 4  & 1024 & 1 & 1 & 128 & 22.7      & 22.5         & 3.88 & 15.2         \\
6   & USEF-TFGridNet & No                                                               & 1  & 4  & 1024 & 4 & 1 & 32  & 22.8      & 22.7         & 3.89 & 14.3         \\
7   & USEF-TFGridNet & No                                                               & 1  & 8  & 512  & 1 & 1 & 128 & 23.0      & 22.8         & 3.90 & 15.2         \\
8   & USEF-TFGridNet & No                                                               & 1  & 8  & 512  & 4 & 1 & 32  & 22.6      & 22.4         & 3.87 & 14.3         \\
9   & USEF-TFGridNet & No                                                               & 1  & 8  & 1024 & 1 & 1 & 128 & 23.0      & 22.8         & 3.90 & 15.2         \\
10  & USEF-TFGridNet & No                                                               & 1  & 8  & 1024 & 4 & 1 & 32  & 22.3      & 22.1         & 3.84 & 14.3         \\ \hline
\end{tabular}
\end{table*}

\section{Results And Discussions}
We present results from \added{three} sets of experiments. \added{The first set of experiments is conducted on the WSJ0-2Mix dataset, the second on the WHAM! and WHAMR! datasets, and the third on the LibriMix dataset and the blind test set from the ICASSP 2023 DNS Challenge.}

\subsection{Results on WSJ0-2mix}
\subsubsection{\added{Comparison Study of USEF-TSE with Speaker Embedding based Models}}
\added{Table~\ref{tab:cmp-cost} presents the results of the USEF-TSE framework and speaker embedding-based models on the WSJ0-2mix dataset, along with their model sizes, training times, and computational costs. By comparing X-SepFormer and Res-SepFormer-v2, it can be observed that although both models have comparable model size, computational cost, and training time, X-SepFormer achieves better performance. This may be attributed to its integration of speaker embeddings at multiple network depths and the use of training strategies specifically designed to reduce speaker confusion errors. By comparing Res-TFGridNet-v2 and X-TF-GridNet, it can be observed that X-TF-GridNet performs better on the WSJ0-2mix dataset. This improvement may be attributed to its multi-task training strategy, which jointly optimizes the separator and the speaker embedding extractor. However, this also results in a longer training time.}

\added{The results in Table~\ref{tab:cmp-cost} show that USEF-SepFormer and USEF-TFGridNet, utilizing the USEF-TSE framework, achieve better SI-SDRi than speaker embedding-based models.  Comparing Res-TFGridNet-v1 and Res-TFGridNet-v2, we observe that while the model size and computational cost of Res-TFGridNet-v2 are significantly higher than those of Res-TFGridNet-v1, its SI-SDRi only improves by 0.4 dB. Furthermore, when comparing Res-TFGridNet-v2 with USEF-TFGridNet, we find that their model sizes and computational costs are comparable, yet USEF-TFGridNet achieves a 3.4 dB higher SI-SDRi than Res-TFGridNet-v2. A similar pattern can be seen when comparing USEF-SepFormer, Res-SepFormer-v1, and Res-SepFormer-v2. These results suggest that the primary reason for the performance improvement is using the USEF-TSE framework rather than the increased model size and computational cost.}
\begin{table*}
\caption{SDRi(dB) and SI-SDRi(dB) of SS and TSE models on the WSJ0-2mix dataset. ‘SS’ and ‘TSE’ denote speech separation and target speaker extraction, respectively. For TSE, we report the average \added{performance} on the two speakers. The average length of the reference speech is 7.3s, which is the same as SpEx+.}
  \label{tab:wsj0-2mix}
    \centering
\begin{tabular}{c|c|c|c|c|c|c|c}
\hline
\textbf{Task}                  & \textbf{Systems}        & \textbf{Domain} & \textbf{Year} & \begin{tabular}[c]{@{}c@{}}\textbf{Use Speaker} \\ \textbf{Embedding?}\end{tabular} & \textbf{Parameter(M)} & \textbf{SDRi(dB)} & \textbf{SI-SDRi(dB)} \\ \hline
\multirow{15}{*}{SS}  & DPCL\cite{hershey2016deep}           & T-F    & 2016 & -                                                                 & 13.6      & -        & 10.8        \\
                      & uPIT\cite{yu2017permutation}           & T-F    & 2017 & -                                                                 & 92.7      & 10.0     & -           \\
                      & DANet\cite{chen2017deep}          & T-F    & 2018 & -                                                                 & 9.1       & 10.8     & 10.4        \\
                      & TasNet\cite{luo2018tasnet}         & Time   & 2018 & -                                                                 & 23.6      & 13.6     & 13.2        \\
                      & Conv-TasNet\cite{luo2019conv}    & Time   & 2019 & -                                                                 & 5.1       & 15.6     & 15.3        \\
                      & Sudo rm -rf\cite{tzinis2020sudo}    & Time   & 2020 & -                                                                 & 2.6       & -        & 18.9        \\
                      & DPRNN\cite{luo2020dual}          & Time   & 2020 & -                                                                 & 2.6       & 19.0     & 18.8        \\
                      & DPTNet\cite{chen2020dual}         & Time   & 2020 & -                                                                 & 2.7       & 20.6     & 20.2        \\
                      & Wavesplit\cite{zeghidour2021wavesplit}      & Time   & 2021 & -                                                                 & 29.0      & 21.2     & 21.0        \\
                      & SepFormer\cite{subakan2021attention}      & Time   & 2021 & -                                                                 & 26.0      & 20.5     & 20.4        \\
                      & QDPN\cite{rixen2022qdpn}           & Time   & 2022 & -                                                                 & 200       & -        & 22.1        \\
                      & TFPSNet\cite{yang2022tfpsnet}        & T-F    & 2022 & -                                                                 & 2.7       & 21.3     & 21.1        \\
                      & TF-GridNet\cite{wang2023tf}     & T-F    & 2022 & -                                                                 & 14.5      & 23.6     & 23.5        \\
                      & MossFormer\cite{zhao2023mossformer}     & Time  & 2023 & -                                                                 & 42.1      & -        & 22.8       \\
                      & MossFormer2\cite{zhao2024mossformer2}    & Time  & 2024 & -                                                                 & 55.7      & -        & 24.1       \\ \hline
\multirow{12}{*}{TSE} & SpeakerBeam\cite{vzmolikova2019speakerbeam}    & T-F    & 2019 & Yes                                                               & -         & 10.9     & -           \\
                      & SpEx\cite{xu2020spex}           & Time   & 2020 & Yes                                                               & 10.8      & 17.0     & 16.6        \\
                      & SpEx+\cite{ge2020spex+}          & Time   & 2020 & Yes                                                               & 11.1      & 17.6     & 17.4        \\
                      & WASE~\cite{hao2021wase}           & Time   & 2021 & Yes                                                               & 7.5       & 17.0     & -           \\
                      & SpEx++\cite{ge2021multi}         & Time   & 2021 & Yes                                                               & -         & 18.4     & 18.0        \\
                      & SpEx$_{pc}$\cite{wang2021neural}         & Time   & 2021 & Yes                                                               & 28.4      & 19.2     & 19.0        \\
                      & X-SepFormer\cite{liu2023x}    & Time   & 2023 & Yes                                                               & -         & 19.7     & 19.1        \\
                      & X-TF-GridNet\cite{hao2024x}   & T-F   & 2023 & Yes                                                               & 12.7       & 21.7     & 20.7        \\
                      & SEF-Net\cite{zeng2023sef}        & Time   & 2023 & No                                                                & 27        & 17.6     & 17.2        \\
                      & VEVE\cite{yang2023target}           & Time   & 2023 & No                                                                & 2.6       & 19.2     & 19.0        \\
                      & SMMA-Net\cite{hu2024smma}       & T-F    & 2024 & No                                                                & 1.6       & 20.6     & 20.4        \\
                      & CIENet-mDPTNet\cite{yang2024target} & T-F    & 2024 & No                                                                & 2.9       & 21.6     & 21.4        \\ \hline
\multirow{2}{*}{TSE}  & USEF-SepFormer & Time   & -    & No                                                                & 19.7      & \textbf{20.3}     & \textbf{19.9}       \\
                      & USEF-TFGridNet & T-F    & -    & No                                                                & 15.2      & \textbf{23.5}     & \textbf{23.3}        \\ \hline
\end{tabular}
\end{table*}
\subsubsection{Ablation Results of USEF-SepFormer With Different Hyper-Parameters}
Table~\ref{tab:abl-usef-sepformer} presents the ablation results of the USEF-SepFormer on WSJ0-2mix, using different model hyperparameters. Critical parameters of the USEF-SepFormer model are detailed in Table~\ref{tab:hyp-usef-sepformer}. Comparing row 3 with row 7 (or row 4 with row 8, row 5 with row 9, and row 6 with row 10), it is evident that the USEF-SepFormer model with 4 attention layers in the CMHA module outperforms the model with only 1 attention layer (\(C_{l} = 4\) vs. \(C_{l} = 1\)). Further, comparing row 7 with row 8 (or rows 9 and 10) shows that the best performance is achieved when the SepFormer backbone module has 8 attention layers and a feed-forward network dimension of 1024 (\(S_{l} = 4\), \(S_{d} = 1024\)). Finally, comparing rows 8 and 10 (or rows 4 and 6) reveals that with \(S_{l} = 4\) and \(S_{d} = 1024\), having 8 attention heads in the CMHA module of USEF-SepFormer yields better performance than having 4 attention heads (SI-SDRi = 19.9 dB vs. SI-SDRi = 19.2 dB). It may be because the CMHA module is consistent with the attention modules used in the separator, which ensures that the frame-level target speaker features and fusion feature reside in a highly similar feature space.

\subsubsection{Ablation Results of USEF-TFGridNet With Different Hyper-Parameters}
Table~\ref{tab:abl-usef-tfgridnet} presents the ablation results of the USEF-TFGridNet on WSJ0-2mix using different model hyper-parameters. The key parameters for the USEF-SepFormer model are detailed in Table~\ref{tab:hyp-usef-tfgridnet}. In USEF-TFGridNet, the self-attention parameters in the separator are consistent with those in TF-GridNet (\(S_{l} = 1\), \(S_{h} = 4\), \(S_{d} = 512\)). Comparing rows 3, 5, and 7, we observe that when the Unfold kernel size is set to 1  (\(K = 1\)), increasing the number of attention heads (\(C_{h}\)) or the dimension of the feed-forward network (FFN) (\(C_{d}\)) in the CMHA module does not enhance the model’s performance. Additionally, comparing rows 3 and 4, as well as rows 7 and 8, shows that when the FFN dimension is 512 (\(C_{d} = 512\)), using Unfold with a kernel size of 4 (\(K = 4\)) leads to decreased performance in the USEF-TFGridNet. Like USEF-Sepformer, USEF-TFGridNet achieves the best results (SI-SDRi = 23.3 dB) when the parameter settings of the CMHA module are consistent with those of the attention modules in the separator.

\subsubsection{Comparison With Previous Models}
Tabel~\ref{tab:wsj0-2mix} compares the performance of our proposed methods with previous speech separation and TSE models on WSJ0-2mix. Our time-domain model, USEF-SepFormer, outperforms the best previous time-domain TSE models using speaker embedding, such as SpEx+, SpEx$_{pc}$, and X-SepFormer, with an SI-SDRi of 19.9 dB compared to 17.4 dB, 19.0 dB, and 19.1 dB, respectively. Additionally, compared to the best time-domain speaker-embedding-free TSE models like SEF-Net and VEVE, USEF-SepFormer achieves state-of-the-art results in both SDRi and SI-SDRi (20.2 dB and 19.9 dB). These results demonstrate that our enhancements to the previous SEF-Net model are practical, significantly improving the performance of time-domain TSE models.
\begin{table*}
\caption{Performance and Computation cost in a comparative study of different feature fusion methods. GMAC/s is computed based on a 1-second mixed speech segment and 8.6-second reference speech.}
  \label{tab:fusion}
  \centering
\begin{tabular}{c|c|c|c|c|c|c}
\hline
\textbf{Model}  & \textbf{Fusion Method} & \textbf{\begin{tabular}[c]{@{}c@{}}SDRi\\ (dB)\end{tabular}} & \textbf{\begin{tabular}[c]{@{}c@{}}SI-SDRi\\ (dB)\end{tabular}} & \textbf{PESQ} & \textbf{\begin{tabular}[c]{@{}c@{}}Parameter\\ (M)\end{tabular}} & \textbf{GMAC/S} \\ \hline
USEF-SepFormer & FiLM                   & 20.3                                                         & 19.9                                                            & 3.67 & 19.7                                                          &  45.8      \\
USEF-SepFormer & Concatenate            & 19.7                                                         & 19.4                                                            & 3.62 & 19.6                                                              & 46.0       \\
USEF-TFGridNet & FiLM                   & 23.5                                                            & 23.3                                                            & 3.92     & 15.4                                                          & 129.0    \\
USEF-TFGridNet & Concatenate            & 23.5                                                         & 23.3                                                            & 3.92 & 15.2                                                          & 125.3       \\ \hline
\end{tabular}
\end{table*}
%
Compared to previous SOTA T-F domain TSE models using speaker embedding, such as SMMA-Net and CIENet-mDPTNet, our proposed T-F domain model USEF-TFGridNet demonstrates superior performance, with an SI-SDRi of 23.3 dB versus 20.4 dB and 21.4 dB, respectively. While the separators in SMMA-Net, CIENet-mDPTNet, and USEF-TFGridNet all process features along the channel dimension, SMMA-Net and CIENet-mDPTNet perform cross-attention on the real and imaginary parts of the reference and mixed speech’s STFT features along the frequency dimension. The resulting features are then concatenated with the mixed speech’s STFT features along the channel dimension before increasing the number of channels. In contrast, our USEF-TFGridNet model first increases the number of channels and then applies cross-attention mechanisms directly along the channel dimension. This consistency in the modeling dimensions between the CMHA and separator modules leads to better performance. Overall, our proposed USEF-TFGridNet outperforms other time or T-F domain TSE models.


%
\begin{figure}[t!]
  \centering
  \includegraphics[width=1.0\linewidth]{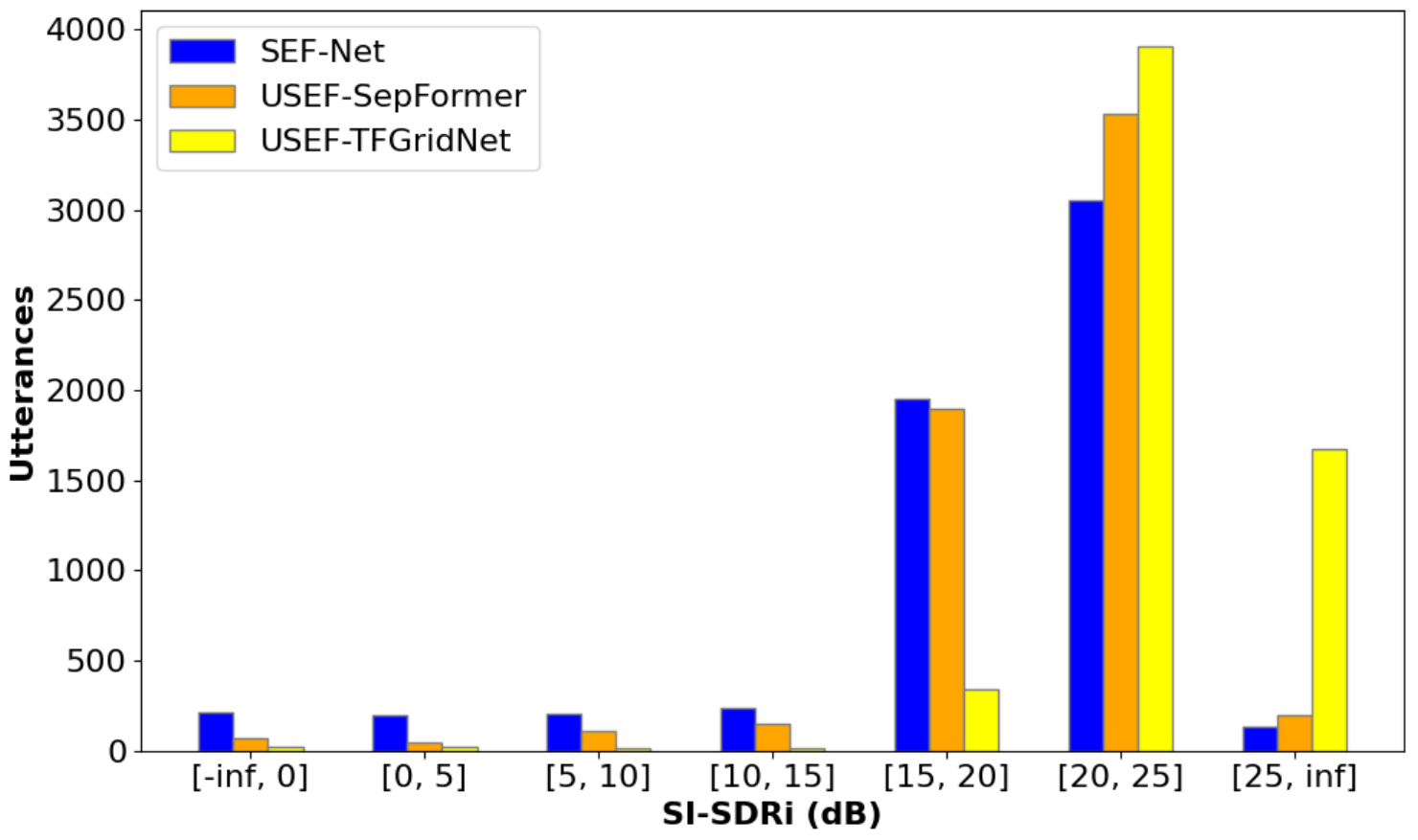}
  \caption{Distributions of the number of test utterances at various SI-SDRi dB ranges.‘Inf’ stands for infinity.}
  \label{fig:dbrange}
\end{figure}
Notably, the backbone networks of X-SepFormer and X-TF-GridNet are identical to those used in our proposed USEF-SepFormer and USEF-TFGridNet. The critical difference lies in the auxiliary features used: X-SepFormer and X-TF-GridNet employ a pre-trained speaker verification model to extract X-vectors as auxiliary features for input into the separator, while USEF-SepFormer and USEF-TFGridNet directly input the acoustic features of the reference speech into the separator. This suggests that, in addition to speaker identity, other aspects of the reference speech, such as contextual information, can also be valuable for accurately extracting the target speaker.

We also compared our proposed models with several mainstream speech separation models on the WSJ0-2mix dataset. Although SS tends to outperform TSE in terms of quality, the former requires prior knowledge of the number of speakers in the mixture and cannot identify the speaker associated with each channel. The current SOTA performance in the SS task is held by the MossFormer2 model, with an SI-SDRi of 24.1 dB.  Previously, the best performance in the target speaker TSE task is achieved by the CIENet-mDPTNet model, with an SI-SDRi of 21.4 dB. Our proposed USEF-TFGridNet has set a new SOTA in the TSE task, achieving an SI-SDRi of 23.2 dB. This represents a 1.8 dB improvement over the CIENet-mDPTNet model, significantly narrowing the performance gap between the TSE and SS tasks from the previous 11.2\% to the current 3.3\% (from 2.7 dB to 0.8 dB).
\begin{table}
\caption{SDR (dB) and PESQ in a comparative study of different and same gender mixed speech. We only report the first speaker in the mixed speech.}
  \label{tab:diff-same}
  \centering
\begin{tabular}{c|cc|cc}
\hline
\multirow{2}{*}{\textbf{Methods}} & \multicolumn{2}{c|}{\textbf{SDR(dB)}} & \multicolumn{2}{c}{\textbf{PESQ}} \\
                                  & \textbf{Diff}   & \textbf{Same}   & \textbf{Diff}   & \textbf{Same}   \\ \hline
Mixture                           & 2.50            & 2.70            & 2.29            & 2.34            \\
SpeakerBeam\cite{vzmolikova2019speakerbeam}                       & 12.01           & 6.87            & 2.82            & 2.43            \\
SpEx\cite{xu2020spex}                              & 19.28           & 14.72           & 3.53            & 3.16            \\
SpEx+\cite{ge2020spex+}                             & 20.08           & 16.77           & 3.62            & 3.34            \\
SEF-Net\cite{zeng2023sef}                           & 20.20           & 18.73           & 3.56            & 3.48            \\
USEF-SepFormer                    & 22.07           & 21.07           & 3.76            & 3.58            \\
USEF-TFGridNet                    & 24.83           & 24.41           & 3.96            & 3.88            \\ \hline
\end{tabular}
\end{table}

\subsubsection{Comparison of Different Fusion Methods}

Table~\ref{tab:fusion} presents the experimental results of USEF-SepFormer and USEF-TFGridNet with different feature fusion methods. Both FiLM and concatenation are effective for feature fusion in the USEF-TSE framework. Notably, FiLM generally outperforms concatenation in USEF-TSE, yielding better results (19.9 dB vs. 19.4 dB and 23.3 dB vs. 23.3 dB). However, in USEF-TFGridNet, while FiLM provides a performance boost, it also introduces additional network structures, slightly increasing the model’s parameters and computational complexity compared to concatenation.

\subsubsection{Comparison of Distributions of the number of test utterances at various dB ranges}
Figure~\ref{fig:dbrange} illustrates the distribution of SI-SDRi for 6,000 target speaker predictions (covering all speakers from 3,000 mixed speech samples). Predictions with an SI-SDRi value below 0 are considered poor cases, and fewer poor cases indicate better model performance. Notably, USEF-SepFormer and USEF-TFGridNet show significantly fewer poor cases compared to SEF-Net. Additionally, USEF-SepFormer and USEF-TFGridNet have a higher concentration of predictions in the 20 to 25 dB range than SEF-Net. The majority of USEF-TFGridNet’s predictions exceed 20 dB, with significantly more predictions above 25 dB compared to USEF-SepFormer and SEF-Net, indicating that USEF-TFGridNet achieves higher-quality target speaker estimation.

\subsubsection{Comparison of Different and Same Gender}
Table~\ref{tab:diff-same} reports the speaker extraction performance of our proposed methods on both different-gender and same-gender mixed speech. To facilitate comparison with other TSE models, we report the experimental results for only the first speaker. It is well-known that separating same-gender mixed speech is more challenging than different-gender speech, resulting in lower SDR and PESQ scores in same-gender scenarios. As shown in the table, both USEF-SepFormer and USEF-TFGridNet outperform SEF-Net in both scenarios. USEF-SepFormer demonstrates a relative improvement over SEF-Net of 22.9\% in SDR and 11.2\% in PESQ for different-gender scenarios, with an even more significant relative improvement of 30.3\% in SDR and 11.5\% in PESQ for same-gender scenarios. Additionally, USEF-TFGridNet notably reduces the SDR gap between same-gender and different-gender scenarios from 1.47 dB to 0.42 dB. This suggests that our proposed model is more effective and consistent in handling same-gender mixed speech scenarios.
\begin{table*}
\caption{SDRi(dB) and SI-SDRi(dB) of SS and TSE models on the WHAM! and WHAMR! datasets. ‘SS’ and ‘TSE’ denote speech separation and target speaker extraction, respectively.}
  \label{tab:wham_whamr}
  \centering
\begin{tabular}{c|c|c|c|cc|cc}
\hline
\multirow{2}{*}{\textbf{Task}} & \multirow{2}{*}{\textbf{Systems}} & \multirow{2}{*}{\textbf{Domain}} & \multirow{2}{*}{\textbf{Params(M)}} & \multicolumn{2}{c|}{\textbf{WHAM!}}      & \multicolumn{2}{c}{\textbf{WHAMR!}}      \\
                               &                                   &                                  &                                     & \textbf{SDRi(dB)} & \textbf{SI-SDRi(dB)} & \textbf{SDRi(dB)} & \textbf{SI-SDRi(dB)} \\ \hline
\multirow{7}{*}{SS}            & Conv-TasNet\cite{luo2019conv}                       & Time                             & 5.1                                 & -                 & 12.7                 & -                 & 8.3                  \\
                               & DPRNN\cite{luo2020dual}                             & Time                             & 2.6                                 & -                 & 13.9                 & -                 & 10.3                 \\
                               & Wavesplit\cite{zeghidour2021wavesplit}                         & Time                             & 29                                  & -                 & 16.0                 & -                 & 13.2                 \\
                               & SepFormer\cite{subakan2021attention}                         & Time                             & 25.7                                & -                 & 16.4                 & -                 & 14.0                 \\
                               & MossFormer\cite{zhao2023mossformer}                        & Time                             & 42.1                                & -                 & 17.3                 & -                  & 16.3                 \\
                               & MossFormer2\cite{zhao2024mossformer2}                       & Time                             & 55.7                                & -                 & 18.1                 & -                  & 17.0
                               \\ 
                               & \added{TF-GridNet\cite{zhao2024mossformer2}}                       & \added{T-F}                             & \added{-}                                & \added{-}                 & \added{-}                 & \added{15.8}                  & \added{17.3} \\
                               \hline
\multirow{5}{*}{TSE}           & SpEx\cite{xu2020spex}                              & Time                             & 10.8                                & 13.0              & 12.2                 & 9.5               & 10.3                 \\
                               & SpEx+\cite{ge2020spex+}                             & Time                             & 11.1                                & 13.6              & 13.1                 & 10.0              & 10.9                 \\
                               & SpEx++\cite{ge2021multi}                            & Time                             & -                                   & 14.7              & 14.3                 & 10.7              & 11.7                 \\
                               & X-TF-GridNet\cite{hao2024x}                      & T-F                              & 12.7                                & 15.7              & 15.3                 & 13.8              & 14.6                 \\
                               & CIENet-mDPTNet\cite{yang2024target}                    & T-F                              & 2.9                                 & 17.0              & 16.6                 & 14.3              & 15.7                 \\ \hline
\multirow{4}{*}{TSE}    & \added{Res-SepFormer-v2}                    & \added{Time}                             & \added{19.7}              & \added{13.7} &\added{12.9}                 & \added{8.1}              & \added{8.6}                  \\       
& USEF-SepFormer                    & Time                             &19.7                                 & 15.5              & 15.1                 & 10.3              & 11.2                 \\
& \added{Res-TFGridNet-v2}                    & \added{T-F}                             &\added{15.2}                                 &\added{16.6}               &\added{15.9}                  & \added{11.6}              & \added{12.0}                 \\
                               & USEF-TFGridNet                    & T-F                              & 15.2                      & \textbf{17.9}     & \textbf{17.6}        & \textbf{14.9}     & \textbf{16.1}        \\ \hline
\end{tabular}
\vspace{-0.4cm}
\end{table*}
\begin{table}
\caption{\added{SDRi (dB) and SI-SDRi (dB) of the TSE models on the LibriMix dataset under the Max mode.}}
  \label{tab:LibriMix}
  \centering
\begin{tabular}{c|cc}
\hline
\textbf{\added{Model}}    & \textbf{\added{SDRi (dB)}} & \textbf{\added{SI-SDRi (dB)}} \\ \hline
\added{SpEx+\cite{ge2020spex+}}             & \added{10.1}              & \added{9.0}                  \\
\added{USED~\cite{Ao_2024}}    & \added{13.2}              & \added{12.7}                 \\ \hline
\added{Res-SepFormer-v2} & \added{4.4}                   & \added{3.6}                      \\
\added{USEF-SepFormer}    & \added{12.0}              & \added{11.4}                 \\
\added{Res-TFGridNet-v2} & \added{12.4}              & \added{12.0}                 \\
\added{USEF-TFGridNet}    & \added{17.4}              & \added{16.8}                 \\ \hline
\end{tabular}
\vspace{-0.4cm}
\end{table}

\subsection{Results on WHAM! and WHAMR!}
Table~\ref{tab:wham_whamr} compares the SDRi and SI-SDRi of our proposed methods with previous SS and TSE models on WHAM! and WHAMR!. USEF-SepFormer outperforms the previous best time-domain TSE model, SpEx++, on WHAM! (15.1 dB vs. 14.3 dB). However, it performs slightly worse on WHAMR! (11.3 dB vs. 11.7 dB). In contrast, T-F domain TSE models like USEF-TFGridNet and CIENet-mDPTNet significantly outperform time-domain models such as SpEx++ and USEF-SepFormer on both datasets. Moreover, the performance of time-domain TSE models degrades more in reverberant environments than T-F domain models. This suggests that T-F domain models may be better suited for TSE in complex scenarios than time-domain models. USEF-TFGridNet achieves SOTA performance on both the WHAM! and WHAMR! datasets. It shows a 5.3\% relative improvement in SDRi and a 6.0\% relative improvement in SI-SDRi over CIENet-mDPTNet on the WHAM! dataset. On the WHAMR! dataset, it demonstrates a 4.2\% relative improvement in SDRi and a 2.5\% relative improvement in SI-SDRi.

We also compared our proposed models with several leading speech SS models on the WHAM! and WHAMR! datasets. MossFormer2, which has achieved SOTA performance in SS tasks, reports an SI-SDRi of 18.1 dB and 17.0 dB on the WHAM! and WHAMR! datasets, respectively. Previously, the SOTA performance for TSE on these datasets is held by the CIENet-mDPTNet model, with SI-SDRi scores of 16.6 dB and 15.7 dB, respectively. Our proposed USEF-TFGridNet sets new state-of-the-art records in TSE tasks, achieving SI-SDRi scores of 17.6 dB and 16.1 dB on the WHAM! and WHAMR! datasets, respectively. Furthermore, USEF-TFGridNet reduces the performance gap between TSE and SS tasks from 8.3\% and 7.6\% to 2.8\% and 5.3\% on the WHAM! and WHAMR! datasets, respectively. \added{We observe that USEF-SepFormer and USEF-TFGridNet perform worse than the original SS models SepFormer and TF-GridNet on the WHAMR! dataset. A possible reason is that compared to the original SepFormer and TF-GridNet, USEF-SepFormer and USEF-TFGridNet are required to reconstruct clean speech for a specific speaker. During this process, speaker confusion can lead to performance degradation. Moreover, the inefficient fusion between the clean reference speech and the reverberant mixed speech further exacerbates the performance drop in reverberant conditions.}
\begin{table}
\caption{\added{DNSMOS of the TSE models on the blind test set from the ICASSP 2023 DNS Challenge. ‘Track 1’ denotes the Headset track. ‘Track 2’ denotes the Speakerphone track. ‘USEF-TFGridNet-EC’ refers to the USEF-TFGridNet model with the Energy Compression (EC) strategy applied.}}
  \label{tab:dns}
  \centering
\resizebox{0.48\textwidth}{!}{
\begin{tabular}{c|ccc|ccc}
\hline
\multirow{2}{*}{\textbf{\added{Model}}} & \multicolumn{3}{c|}{\textbf{\added{Track 1}}} & \multicolumn{3}{c}{\textbf{\added{Track 2}}} \\
                                & \textbf{\added{SIG}}  & \textbf{\added{BAK}} & \textbf{\added{OVRL}} & \textbf{\added{SIG}}   & \textbf{\added{BAK}}   & \textbf{\added{OVRL}}  \\ \hline
\added{Noisy}                           & \added{4.13}         & \added{2.42}        & \added{2.73}        & \added{4.02}          & \added{2.20}          & \added{2.51}         \\ \hline
\added{SpEx+\cite{ge2020spex+}}                           & \added{2.73}         & \added{3.87}        & \added{2.45}        & \added{2.43}          & \added{3.82}          & \added{2.22}         \\
\added{CIENet-mDPTNet\cite{yang2024target}}                  & \added{3.74}         & \added{3.79}        & \added{3.24}        & \added{3.63}          & \added{3.66}          & \added{3.10}         \\ \hline
\added{Res-SepFormer-v2}               & \added{2.86}         & \added{2.43}        & \added{2.06}        & \added{2.59}          & \added{2.30}          & \added{1.82}         \\
\added{USEF-SepFormer}                  & \added{3.33}         & \added{2.88}        & \added{2.57}        & \added{3.01}          & \added{2.69}          & \added{2.26}         \\
\added{Res-TFGridNet-v2}               & \added{3.43}         & \added{3.23}        & \added{2.80}        & \added{3.31}          & \added{3.21}          & \added{2.70}         \\
\added{USEF-TFGridNet}                  & \added{3.63}         & \added{3.64}        & \added{3.12}        & \added{3.53}          & \added{3.51}          & \added{2.96}         \\
\added{USEF-TFGridNet-EC}                  & \added{3.76}         & \added{3.65}        & \added{3.22}        & \added{3.64}          & \added{3.62}          & \added{3.10}         \\ \hline
\end{tabular}}
\vspace{-0.4cm}
\end{table}
\subsection{\added{Results on LibriMix and DNS Challenge Dataset}}
\added{Table~\ref{tab:LibriMix} presents the results of the proposed TSE models on the LibriMix dataset under the max mode. The results in Table~\ref{tab:LibriMix} show that USEF-SepFormer and USEF-TFGridNet, based on the USEF-TSE framework, achieve significantly higher SI-SDRi compared to Res-SepFormer-v2 and Res-TFGridNet-v2 (16.8 dB, 11.4 dB vs. 12.0 dB, 3.6 dB).
USED is a joint speaker extraction and diarization system that achieves strong performance on the LibriMix dataset. However, our model outperforms USED in terms of SI-SDRi (16.8 dB vs. 12.7 dB). It demonstrates that our proposed USEF-TSE framework performs well on more diverse datasets. Additionally, we observe that Res-SepFormer-v2 achieves the lowest SI-SDRi. This is because, compared to WSJ0-2mix, LibriMix involves more speakers and, in the max mode, includes segments with partial overlap. The test utterances contain 2-speaker or 3-speaker mixtures and include audio segments where only the target or interfering speaker is active. In such cases, the possibility of speaker confusion~\cite{zhao2022target} increases, making the TSE task more challenging. Additionally, unlike the SpEx+ model, the Res-SepFormer-v2 model does not adopt a multi-task training strategy. Instead, it directly interacts with the speaker embedding and the time-domain features of the mixed speech. This mismatch in feature domains further increases the risk of speaker confusion, ultimately leading to a significant drop in SI-SDRi performance.}  

\added{Table~\ref{tab:dns} presents the results of TSE models on the blind test set from the ICASSP 2023 Challenge. In this set of experiments, we use WHAMR! as the training set to validate the performance of our proposed model on out-of-domain data. Table~\ref{tab:dns} shows that on the headset and speakerphone tracks, USEF-SepFormer achieves results comparable to SpEx+ (OVRL: 2.57, 2.26 vs. 2.45, 2.22), and USEF-TFGridNet outperforms SpEx+(OVRL: 2.57, 2.26 vs. 3.12, 2.96). This indicates that our proposed USEF-TSE framework effectively performs generalization on out-of-domain data. However, the results of USEF-TFGridNet are slightly worse than those of CIENet-mDPTNet. A possible reason is that CIENet-mDPTNet employs an energy compression strategy~\cite{li2021importance}. By applying nonlinear compression to amplitude or power, the dynamic range of the data is reduced, which enhances the model’s robustness to background noise and improves its stability and generalization. To improve the model’s generalization performance, we applied the energy compression strategy to the USEF-TFGridNet, resulting in the USEF-TFGridNet-EC model. Table~\ref{tab:dns} shows that USEF-TFGridNet-EC and CIENet-mDPTNet achieve comparable performance on the Speakerphone and Headset tracks, demonstrating strong generalization ability on out-of-domain data.}
 

\section{Conclusion}
In this paper, we propose a Universal Speaker Embedding-Free Target Speaker Extraction (USEF-TSE) framework for monaural target speaker extraction. To validate the effectiveness of USEF-TSE, we integrate it with SepFormer and TF-GridNet to propose two target speaker extraction models: USEF-SepFormer and USEF-TFGridNet. Experimental results demonstrate that USEF-SepFormer and USEF-TFGridNet perform better than other time-domain and time-frequency domain target speaker extraction models. Specifically, USEF-TFGridNet achieves state-of-the-art SI-SDRi scores of 23.3 dB on the WSJ0-2mix dataset, 17.6 dB on the WHAM! dataset, and 16.1 dB on the WHAMR! dataset for the TSE task. Furthermore, USEF-TFGridNet significantly narrows the performance gap between TSE and speech separation tasks, highlighting the robustness and effectiveness of our USEF-TSE framework under noisy and reverberant conditions. \added{The results on the LibriMix and the blind test set of the ICASSP 2023 DNS Challenge demonstrate that the USEF-TSE performs well on more diverse and out-of-domain data.} Future work will focus on addressing the limitations of the USEF-TSE framework, \added{such as further improving its generalization performance on out-of-domain data,} which represents a promising research direction.

\section{Acknowledgments}
This research is funded in part by the National Natural Science Foundation of China (62171207), Science and Technology Program of Suzhou City(SYC2022051) and Guangdong Science and Technology Plan (2023A1111120012). Many thanks for the computational resource provided by the Advanced Computing East China Sub-Center.


\bibliographystyle{IEEEtran}
\bibliography{refs}
%

\begin{IEEEbiography}[{\includegraphics[width=1in,height=1.25in,clip,keepaspectratio]{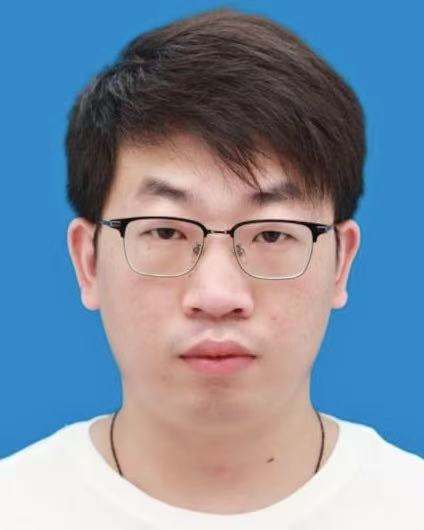}}]{Bang Zeng} received the B.Eng degree from Wuhan University of Science and Technology, China and M.Eng degree from Wuhan University, China, in 2017 and 2021, respectively. He is currently a Ph.D Student in School of Computer Science, Wuhan University, China. His research interests include speech separation, target speaker extraction and personal voice activity detection.
\end{IEEEbiography}
\begin{IEEEbiography}[{\includegraphics[width=1in,height=1.25in,clip,keepaspectratio]{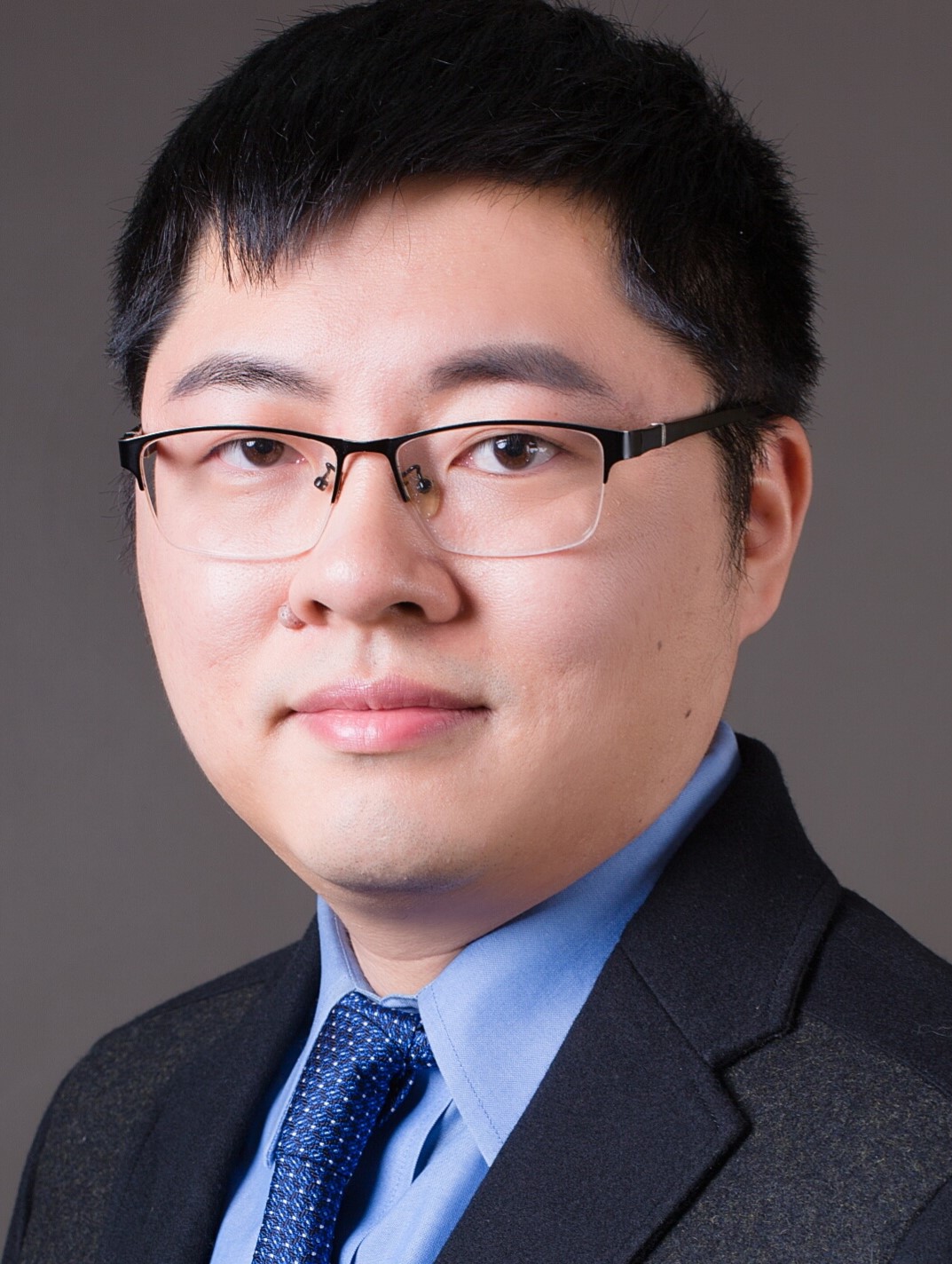}}]{Ming Li}
received his Ph.D. in Electrical Engineering from University of Southern California in 2013. He is currently a Professor of Electronical and Computer Engineering at Duke Kunshan University. He is also an Adjunct Professor at School of Computer Science in Wuhan University. His research interests are in the areas of audio, speech and language processing as well as multimodal behavior signal processing. He has published more than 200 papers and served as the member of IEEE speech and language technical committee, APSIPA speech and language processing technical committee, the editorial board member of the IEEE/ACM Transactions on Audio, Speech, and Language Processing and Computer Speech \& Language. He is an area chair at Interspeech 2016, 2018, 2020 and 2024, 2025 as well as the technical program co-chair of Odyssey 2022 and ASRU 2023. Works co-authored with his colleagues have won first prize awards at Interspeech Computational Paralinguistic Challenges 2011, 2012 and 2019, ASRU 2019 MGB-5 ADI Challenge, Interspeech 2020 and 2021 Fearless Steps Challenges, VoxSRC 2021, 2022 and 2023 Challenges, ICASSP 2022 M2MeT Challenge, IJCAI 2023 ADD challenge, ICME 2024 ChatCLR challenge and Interspeech 2024 AVSE challenge. He received the IBM faculty award in 2016, the ISCA Computer Speech and Language 5-years best journal paper award in 2018 and the youth achievement award of outstanding scientific research achievements of Chinese higher education in 2020.
\end{IEEEbiography}

\end{document}